\documentclass[aps,twocolumn,preprintnumbers,amsmath,amssymb,nofootinbib,superscriptaddress,notitlepage]{revtex4-1}

\usepackage{epsfig}
\usepackage[utf8]{inputenc}

\usepackage{color}
 
\usepackage{hyperref}
\usepackage{multirow}

\usepackage{tabularx}

\begin{document}

\title{Molecular charmed baryons and pentaquarks from light-meson exchange saturation}

\author{Mao-Jun Yan}
\affiliation{CAS Key Laboratory of Theoretical Physics, 
  Institute of Theoretical Physics, \\
  Chinese Academy of Sciences, Beijing 100190, China}

\author{Fang-Zheng Peng}
\affiliation{School of Physics, Beihang University, Beijing 100191, China} 

\author{Manuel Pavon Valderrama}\email{mpavon@buaa.edu.cn}
\affiliation{School of Physics, Beihang University, Beijing 100191, China} 

\date{\today}
%\date{\today}

\begin{abstract} 
  \rule{0ex}{3ex}
  The spectrum of the $c qq$ baryons contains a few states whose nature
  is not clearly a three-quark composite and which might have
  a sizable baryon-meson component.
  Examples include the $\Sigma_c(2800)$ or the $\Lambda_c(2940)$.
  Here we explore the spectrum of two-body systems composed of a light, octet
  baryon and a charmed meson (or antimeson) within a simple contact-range theory
  in which the couplings are saturated by light-meson exchanges.
  This results in the prediction of a series of composite anticharmed
  pentaquarks ($\bar{c} q qqq $) and singly-charmed baryons ($c \bar{q} qqq $).
  Among the later we find $J=\tfrac{1}{2}$ $\Xi D$ and $J=\tfrac{3}{2}$
  $\Xi D^*$ bound states with masses matching those of the recently
  observed $\Omega_c(3185)$ and $\Omega_c(3327)$ baryons.
\end{abstract}

\maketitle

\section{Introduction}

The discovery of a plethora of new heavy hadrons in experimental facilities
calls for their theoretical description and classification~\cite{Guo:2017jvc,Olsen:2017bmm,Brambilla:2019esw,Chen:2022asf}.
While a large number of them are standard three-quark baryons or
quark-antiquark mesons, others do not easily fit into
this explanation and are suspected to be exotic.
If we consider charmed baryons, a few of them do not conform to
the theoretical expectations for $cqq$ states.
For instance, the mass of the $\Lambda_c(2940)$~\cite{BaBar:2006itc,Belle:2006xni,LHCb:2017jym}
is somewhat large compared with quark model predictions~\cite{Capstick:1986ter,Ebert:2011kk,Chen:2014nyo,Lu:2016ctt} and
is really close to the $D^{*0} p$ threshold,
two factors which have in turn prompted its molecular
interpretation~\cite{He:2006is,He:2010zq,Zhang:2012jk,Ortega:2012cx,Wang:2020dhf,Sakai:2020psu}.
The case of the $\Sigma_c(2800)$~\cite{Belle:2004zjl} is similar~\cite{Haidenbauer:2010ch,Sakai:2020psu}
and there are a few excited $\Xi_c$ states (e.g. the $\Xi_c(3055)$
and $\Xi_c(3123)$~\cite{BaBar:2007zjt}) which might also be amenable
for a bound state explanation owing to their closeness to meson-baryon
thresholds (though the most common theoretical explanation of
the $\Xi_c(3055)$ and $\Xi_c(3123)$ is that they are compact
hadrons~\cite{Liu:2007ge,Zhao:2016qmh,Chen:2017aqm,Wang:2017vtv,Li:2022xtj}).
The $\Omega_c(3185)$ and $\Omega_c(3327)$ states recently observed by
the LHCb~\cite{LHCb:2023rtu} might provide another example, being
really close to the $\Xi D$ and $\Xi D^*$ thresholds (check
for instance Ref.~\cite{Feng:2023ixl} for a molecular explanation
of their decays or Refs.~\cite{Luo:2023sra,Karliner:2023okv}
for their interpretations within the quark model).
Much of the theoretical speculations are driven by the lack of
detailed experimental information about these charmed baryons.
Often, we do not even know their quantum numbers or whether a particular
charmed, non-strange baryon is a $\Lambda_c$ or a $\Sigma_c$
(i.e. the case of the $\Lambda_c(2765)$, which is considered to be
a $\Lambda_c$ in Ref.~\cite{Belle:2019bab}, but it could also
be a $\Sigma_c$ or a superposition of $\Lambda_c$
and $\Sigma_c$~\cite{ParticleDataGroup:2022pth}).

In view of the previous situation, the investigation of the bound state
spectrum of a light baryon and a charmed meson is well justified.
Identifying in which configurations to expect molecular charmed baryons
could improve our priors when confronted with a new experimental
discovery or our posteriors when analyzing previous observations.
To deal with the spectroscopy of these states, we formulate here
a contact-range theory where the couplings are saturated from
light-meson exchanges in the line of what we previously did
in Refs.~\cite{Peng:2020xrf,Peng:2021hkr}.
This approach is indeed able to reproduce a few hadrons which are often
though to be exotic, such as the $X(3872)$~\cite{Choi:2003ue} or
the $Y(4230)$~\cite{Aubert:2005rm}, and in the present manuscript
we will particularize it to the case at hand
(i.e. charmed meson and light baryon).

Regarding the aforementioned $\Sigma_c(2800)$ and $\Lambda_c(2940)$, it turns
out that they might be identified with two of the most attractive
$ND^{(*)}$ configurations within our model, giving support to the idea that
they might be molecular after all~\cite{He:2006is,He:2010zq,Zhang:2012jk,Ortega:2012cx,Wang:2020dhf,Sakai:2020psu,Haidenbauer:2010ch},
though more so for the $\Lambda_c(2940)$ than for the $\Sigma_c(2800)$.
If we turn our attention to the $\Omega_c(3185)$ and $\Omega_c(3327)$,
their masses are easily reproduced if they are $\Xi D$ and $\Xi D^*$
states with spin $J=\tfrac{1}{2}$ and $\tfrac{3}{2}$,
respectively (in the latter case coinciding with the preferred spin
of the $\Omega_c(3327)$ according to Ref.~\cite{Feng:2023ixl}).
Yet, besides these candidates, we are able to identify other attractive
configurations that may lead to a molecular singly charmed baryon
or anticharmed pentaquark (though in this later case
there are no experimental candidates).

\section{Saturation of the contact-range couplings}

We describe the charmed meson - octet baryon interaction
with a contact-range theory of the type
\begin{eqnarray}
  V_C(\vec{q}\,) = C_0 + C_1 \, \vec{\sigma}_{L1} \cdot \vec{\sigma}_{L2} \, ,
\end{eqnarray}
where $C_0$ and $C_1$ are coupling constants, $\vec{\sigma}_{L1}$ and
$\vec{\sigma}_{L2}$ are the light-spin operators for the meson and baryon,
respectively, where for the octet baryon the light spin is identical
to the total spin (as it does not contain a heavy quark), and
$\vec{q}$ is the momentum exchanged between the particles.

This description is valid provided the following conditions are met:
(i) the typical momentum of the particles is considerably smaller
than the mass of the scalar and vector mesons
($\sigma$, $\rho$, $\omega$), and (ii) pion exchanges are perturbative.
We remind that this potential is singular though (it corresponds to
a Dirac-delta in r-space)
and will have to be regularized with a regularization scale or
cutoff $\Lambda$, as we will explain later.

To determine the value of the couplings $C_0$ and $C_1$
we exploit the first of the previous conditions ---
that $|\vec{q}\,| < m_{\sigma}$, $m_{\rho}$, $m_{\omega}$ --- 
and combine it with a specific choice of the regularization scale $\Lambda$.
At low enough momenta the finite-range potential generated by the exchange of
a meson can be effectively approximated by a contact-range potential.
If the regularization scale is of the order of the mass of the aforementioned
meson, i.e. $\Lambda \sim m_{\sigma}$, $m_{\rho}$, $m_{\omega}$, the value of
the contact-range couplings will be saturated by
light-meson exchanges~\cite{Ecker:1988te,Epelbaum:2001fm}.
The scalar meson, which generates the potential
\begin{eqnarray}
  V_S(\vec{q}) &=& - \frac{g_{S1} g_{S2}}{m_S^2 + {\vec{q}\,}^2} \, , 
\end{eqnarray}
will contribute to the couplings $C_0$ and $C_1$ as follows
\begin{eqnarray}
  C_0^S(\Lambda \sim m_S) &\propto& - \frac{g_{S1} g_{S2}}{m_S^2} \, , \\
  C_1^S(\Lambda \sim m_S) &\propto& 0 \, ,
\end{eqnarray}
where $g_{S1}$, $g_{S2}$ are the scalar couplings of hadron $1$ and $2$ and
$m_S$ the mass of the scalar meson.
For the vector mesons, the potential reads
\begin{eqnarray}
  V_V(\vec{q}) &=&
  \frac{g_{V1} g_{V2}}{m_V^2 + {\vec{q}\,}^2}
  + \frac{f_{V1} f_{V2}}{6 M^2}\,\frac{m_V^2}{m_V^2 + {\vec{q}\,}^2}\,
  {\vec{\sigma}}_{L1} \cdot {\vec{\sigma}}_{L2} \nonumber \\
  &+& \dots \, ,
  \label{eq:V-vector}
\end{eqnarray}
where the dots indicate either higher partial wave
operators or Dirac-delta contributions.
This leads to the saturated couplings
\begin{eqnarray}
  C_0^V(\Lambda \sim m_V) &\propto& \frac{g_{V1} g_{V2}}{m_V^2} \, , \\
  C_1^V(\Lambda \sim m_V) &\propto& \frac{f_{V1} f_{V2}}{6 M^2} \, , 
\end{eqnarray}
where we have obviated isospin or flavor factors for simplicity and
with $g_{V1}$, $g_{V2}$ the electric-like couplings, $f_{V1}$, $f_{V2}$
the magnetic-like ones, $m_V$ the mass of the vector meson and $M$
a scaling mass which is often taken to be the nucleon mass
($M = m_N$, with $m_N \approx 940\,{\rm MeV}$).
Here we notice that the higher partial wave operators do not contribute to
the saturation of the S-wave couplings, while the Dirac-delta contributions
are regularized by the finite size of hadrons $1$ and $2$ and only
contribute to the saturation of the couplings at the regularization
scale $\Lambda \sim M_H$, with $M_H$ the characteristic momentum
scale of the finite size effects for a hadron $H$.
In general $M_H \gg m_V$, which is why we ignore the Dirac-delta
contributions~\cite{Peng:2020xrf}.

At this point we encounter a problem: saturation is expected to work for a
regularization scale similar to the mass of the light-meson being exchanged,
yet the masses of the scalar and vector mesons are different.
This means that there is a small mismatch in the ideal saturation scale for
scalar ($\Lambda \sim m_S$) and vector ($\Lambda \sim m_V$) mesons.
This is however easily solvable from the renormalization group (RG) evolution 
of the saturated couplings, which can be derived from the condition
that the matrix elements of the contact-range potential are
independent of the cutoff~\cite{PavonValderrama:2014zeq}
\begin{eqnarray}
  \frac{d}{d \Lambda}\,\langle \Psi | V_C(\Lambda) | \Psi \rangle = 0 \, .
\end{eqnarray}
If the wave function has a power-law behavior $\Psi(r) \sim r^{\alpha/2}$
at distances $r \sim 1/\Lambda$, the RG equation above leads to
\begin{eqnarray}
  \frac{C(\Lambda_1)}{\Lambda_1^{\alpha}} = \frac{C(\Lambda_2)}{\Lambda_2^{\alpha}}
  \, ,
\end{eqnarray}
from which we can combine the scalar and vector meson contributions as
\begin{eqnarray}
  C(m_V) = C^V(m_V) +
  {\left( \frac{m_V}{m_S} \right)}^{\alpha}\,C^S(m_S) \, .
\end{eqnarray}
The intuitive meaning of this equation is that the relative strength of
the contribution of a lighter meson scales as $1/m^{2+\alpha}$ (instead of
$1/m^2$ if we do not consider their RG evolution).
For the exponent $\alpha$ we use the semi-classical approximation
together with the Langer correction~\cite{Langer:1937qr},
leading to $\Psi(r) \sim \sqrt{r}$ or $\alpha = 1$.

Finally,
if we plug in the expected values of the coupling constants from saturation
we end up with 
\begin{eqnarray}
  C^{\rm sat}(m_V) &\propto&
  \frac{g_{\rho 1} g_{\rho 2}}{m_V^2}\,\left( 1 + \kappa_{\rho 1} \kappa_{\rho 2}\,\frac{m_V^2}{6 M^2}\,\hat{S}_{L12} \right)\,\hat{T}_{12} \nonumber \\ &+& 
  \frac{g_{\omega 1} g_{\omega 2}}{m_{V}^2}\,\left( 1 + \kappa_{\omega 1} \kappa_{\omega 2}\,\frac{m_V^2}{6 M^2}\,\hat{S}_{L12} \right)\,\zeta \nonumber
  \\ &+&
  \left( \frac{m_V}{m_\phi}\right)\,\frac{g_{\phi 1} g_{\phi 2}}{m_{\phi}^2}\,\left( 1 + \kappa_{\phi 1} \kappa_{\phi 2}\,\frac{m_\phi^2}{6 M^2}\,\hat{S}_{L12} \right)\,\zeta \nonumber
  \\ &-& 
  \left( \frac{m_V}{m_S}\right)\,\frac{g_{S1} g_{S2}}{m_S^2} \, ,
  \label{eq:C-sat}
\end{eqnarray}
where we have now included isospin factors
($\hat{T}_{12} = \hat{\vec{T}}_1 \cdot \hat{\vec{T}}_2$,
with $\hat{\vec{T}} = \vec{T} / T$ a normalized isospin operator and
$T$ the isospin of the particle),
defined $\hat{S}_{L12} = \vec{\sigma}_{L1} \cdot \vec{\sigma}_{L2}$
and taken into account that $\alpha = 1$.
In the previous equation we use the decomposition $f_V = \kappa_V \, g_V$
for the magnetic-like couplings and introduce the G-parity sign $\zeta$,
which is $\zeta = +1$ or $-1$ for molecular anticharmed pentaquarks
and charmed baryons, respectively.
The $\rho$ and $\omega$ contributions are kept separate
because for the nucleon we have $g_{\rho} \neq g_{\omega}$.
For the masses of the vector mesons we take
$m_V = (m_\rho + m_{\omega})/2 = 775\,{\rm MeV}$ for $V = \rho, \omega$
(i.e. the average of the $\rho$ and
$\omega$ masses) and $m_{\phi} = 1020\,{\rm MeV}$.
The only thing left is the proportionality constant, which can be determined
from the condition of reproducing the binding energy of a known molecular
candidate.

\begin{table}
  \centering
  \begin{tabular}{|c|c|ccc|ccc|}
    \hline \hline
    Hadron & $g_{\sigma}$
    & $g_{\rho}$ & $g_{\omega}$ & $g_{\phi}$
    & $\kappa_{\rho}$ & $\kappa_{\omega}$ & $\kappa_{\phi}$ \\
    \hline  \hline
    $D$, $D^*$ & $\frac{1}{3}\,g_S$ & $g_V$ & $g_V$ & $0$ &
    $\frac{3}{2} \mu_u$ & $\frac{3}{2} \mu_u$ & $0$  \\
    $D_s$, $D_s^*$ & $\frac{1}{3}\,g_S$ & $0$ & $0$ & $\sqrt{2} g_V$ &
    $0$ & $0$ & $-3 \mu_s$ \\
    \hline \hline
    $N$ & $g_S$ & $g_V$ & $3 g_V$ & $0$ & $\frac{5}{2}\,\mu_u$
    & $\frac{1}{2}\,\mu_u$ & 0 \\
    $\Lambda$ & $0.75\,g_S$ & $0$ & $2 g_V$ & $\sqrt{2} g_V$ & $0$
    & $0$ & $-3\,\mu_s$ \\
    $\Sigma$ & $g_S$ & $2 g_V$ & $2 g_V$ & $\sqrt{2} g_V$ & $\mu_u$
    & $\mu_u$ & $-\mu_s$ \\
    $\Xi$ & $g_S$ & $g_V$ & $g_V$ & $2 \sqrt{2} g_V$ & $-\frac{1}{2}\,\mu_u$
    & $-\frac{1}{2}\,\mu_u$ & $2 \mu_s$ \\
    \hline \hline
  \end{tabular}
  \caption{
    Choice of couplings for light-meson exchange saturation in this work.
    For their concrete values we take $g_S = 10.2$, $g_V = 2.9$,
    $\mu_u = 1.9$ and $\mu_s = -0.6$.
  }
  \label{tab:couplings}
\end{table}

\section{Qualitative features of the spectrum}

From the previous formalism we can already determine the qualitative
characteristics of the two-body light baryon and charmed (anti)meson
bound state spectrum.

First, we need the couplings of the scalar and vector mesons
to the light baryons and charmed mesons, for which
we will refer to Table \ref{tab:couplings}.
For the vector mesons ($\rho$, $\omega$ and $\phi$) we have simply made use
of the mixing of these mesons with the electromagnetic current (vector meson
dominance~~\cite{Sakurai:1960ju,Kawarabayashi:1966kd,Riazuddin:1966sw})
as a way to determine the $g_V$ and $\kappa_V$ (E0 and M1)
couplings: we can match $g_V$ and $\kappa_V$ to the charge and
magnetic moment of the particular hadron we are interested in.
The $\kappa_V$ couplings are written in terms of the magnetic moments
of the constituent quarks, $\mu_q$, in units of the nuclear magneton
(we take $\mu_u = 1.9\,{\mu_N}$, $\mu_d = -\mu_u/2 $,
$\mu_s = -0.6\,{\mu_N}$ with $\mu_N$
the nuclear magneton).
For the scalar meson the linear sigma model~\cite{GellMann:1960np} predicts
$g_S = \sqrt{2} m_N / f_{\pi} \simeq 10.2$ for the nucleon,
where $m_N$ is the nucleon mass and $f_{\pi} \simeq 132\,{\rm MeV}$
the pion weak decay constant.
For the charmed meson, which contains one light-quark instead of three,
we assume the quark model~\cite{Riska:2000gd} relation
$g_{S qq} = g_{S} / 3$, i.e. that the coupling of
the sigma is proportional to the number of light-quarks within the hadron.
In the strange sector we will assume that the coupling of the scalar
to the $s$ quark is approximately the same as to the $u$ and $d$
quarks: $g_{S uu} = g_{S dd} = g_{S ss}$.
This assumption works well when comparing the $D\bar{D}$ and $D_s \bar{D}_s$
systems predicted in the lattice and for the $27$-plet dibaryons
(i.e. the $NN$, $\Sigma N$, $\Sigma \Sigma$, $\Sigma \Xi$ and
$\Xi \Xi$ in the ${}^1S_0$ partial wave and in their respective
maximum isospin configurations).
The only exception to this rule will be the $\Lambda$ hyperon, for which
a coupling $g_{S \Lambda \Lambda} \simeq 0.75\,g_S$ is necessary for
reproducing the $N \Lambda$ and $\Lambda \Lambda$ scattering
lengths correctly.
A more complete explanation of our choice can be found
in Appendix~\ref{app:couplings}.

Second, for simplicity in the discussion that follows we will use
the SU(3)-symmetric limit of the vector meson masses and
the previous couplings.
That is, now we will assume $m_{\rho} = m_{\omega} = m_{K^*} = m_{\phi}$,
$\mu_s = -\mu_u/2$ and $g_{S \Lambda \Lambda} = g_S$.
In contrast, for the actual quantitative predictions of the next section,
we will use the values of Table \ref{tab:couplings} and the vector
meson masses specified below Eq.~(\ref{eq:C-sat}).

Third, the light baryons and charmed mesons belong to the $8$ and $\bar{3}$
representations of SU(3)-flavor.
Conversely, the two-hadron interaction can be decomposed in a sum of
contributions from different irreducible representations of SU(3):
\begin{eqnarray}
  V_C = \sum_{R} \lambda^{R} V_C^{R} \, ,
\end{eqnarray}
where $R$ indicates a particular representation and $\lambda^R$
is a numerical factor (actually, the square of the relevant
SU(3) Clebsch-Gordan coefficient, which we take
from~\cite{Kaeding:1995vq}).
For the scalar meson contribution, the decomposition will be trivial
\begin{eqnarray}
  C_S^{R} = - \frac{1}{3}\,\frac{g_S^2}{m_S^2} \, ,
\end{eqnarray}
independently of the representation $R$.

For the vector mesons the decomposition is not trivial, but it is still
straightforward.
If we consider the baryon - charmed meson two-body system, the SU(3)
decomposition is $8 \otimes \bar{3} = \bar{3} \oplus 6 \oplus \bar{15}$.
The electric-type vector meson contributions are
\begin{eqnarray}
  C_{V0}^{\bar{3}} &=& - 8 \frac{g_V^2}{m_V^2} \, , \\
  C_{V0}^{6} &=& - 4 \frac{g_V^2}{m_V^2} \, , \\
  C_{V0}^{\bar{15}} &=& 0 \, ,
\end{eqnarray}
while the magnetic-type ones are
\begin{eqnarray}
  C_{V1}^{\bar{3}} &=& - 8\,g_V^2\,\frac{m_V^2}{6 M^2}\,\kappa_{q}^2 \, , \\
  C_{V1}^{6} &=& + \frac{4}{3}\,g_V^2\,\frac{m_V^2}{6 M^2}\,\kappa_{q}^2 \, , \\
  C_{V1}^{\bar{15}} &=& 0 \, ,
\end{eqnarray}
where $\kappa_q = \frac{3}{2}\,(\mu_u / \mu_N)$, i.e. the value of $\kappa_V$
for a light-quark in the SU(3)-symmetric limit.
If we consider the baryon - charmed antimeson two-body system instead, the SU(3)
decomposition is $8 \otimes {3} = {3} \oplus \bar{6} \oplus {15}$.
In this case, the E0 vector meson contributions are
\begin{eqnarray}
  C_{V0}^{{3}} &=& - 4 \frac{g_V^2}{m_V^2} \, , \\
  C_{V0}^{\bar{6}} &=& 0 \, , \\
  C_{V0}^{{15}} &=& + 4 \frac{g_V^2}{m_V^2} \, , 
\end{eqnarray}
while the M1 are
\begin{eqnarray}
  C_{V1}^{{3}} &=& 0 \, , \\
  C_{V1}^{\bar{6}} &=& -4\,g_V^2\,\frac{m_V^2}{6 M^2}\,\kappa_q^2 \, , \\
  C_{V1}^{{15}} &=& + \frac{8}{3}\,g_V^2\,\frac{m_V}{6 M^2}\,\kappa_q^2 \, .
\end{eqnarray}

The SU(3) decomposition of the light baryon and charmed (anti)meson potential
is shown in Tables \ref{tab:pot-baryon-meson} and
\ref{tab:pot-baryon-antimeson}.
While the strength of scalar meson exchange is the same for all the baryon-meson
molecules in the SU(3) symmetric limit, this is not the case for
vector meson exchange, which is the factor deciding what are
the most attractive molecules.
If we consider the baryon-meson case, the total strength of the central and
spin-spin pieces of vector meson exchange is shown
in Table \ref{tab:pot-baryon-meson}.
For the molecules involving the $D$ and $D_s$ pseudoscalar charmed mesons
the spin-spin interaction does not contribute and, provided all
configurations are attractive enough to bind, we will expect
the following hierarchy for the binding energies
\begin{eqnarray}
  && B_{\rm mol}(N D(0), \Sigma D(\tfrac{1}{2})) \nonumber \\
  && \quad > B_{\rm mol}(\Xi D, \Xi D_s) \nonumber \\
  && \quad > B_{\rm mol}(\Lambda D_s, ND(1), \Lambda D, \Sigma D_s) \nonumber \\
  && \quad > B_{\rm mol}(ND_s, \Sigma D(\tfrac{3}{2}), \Xi D(1)) \, ,
\end{eqnarray}
where $B_{\rm mol}$ if defined as positive (such that the mass of a
two-hadron bound state is given by $M = m_1 + m_2 - B_{\rm mol}$,
with $m_1$, $m_2$ the masses of the hadrons) and
the number in parentheses refers to the isospin of a
given molecule (if there is more than one isospin configuration).
If we change the pseudoscalar charmed mesons by antimesons,
the hierarchy will be instead
\begin{eqnarray}
  && B_{\rm mol}(\Sigma \bar{D}(\tfrac{1}{2}), \Xi \bar{D}(0)) \nonumber \\
  && \quad > B_{\rm mol}(N\bar{D}(0), N\bar{D}_s) \nonumber \\
  && \quad > B_{\rm mol}(\Lambda \bar{D}, \Lambda \bar{D}_s, \Xi \bar{D}(1),
  \Sigma \bar{D}_s) \nonumber \\
  && \quad > B_{\rm mol}(N\bar{D}(1), \Sigma \bar{D}(\tfrac{3}{2}), \Xi \bar{D}_s) \, ,
\end{eqnarray}
though it should be noted that the molecules with charmed antimesons are
in general less attractive than the ones containing charmed mesons,
owing to the sign of $\omega$ and $\phi$ exchange.

For the molecules containing a $D^*$($\bar{D}^*$) or $D_s^*$($\bar{D}_s^*$)
vector charmed (anti)meson, the spin-spin interaction generates a hyperfine
splitting between the $J=\tfrac{1}{2}$ and $\tfrac{3}{2}$ configurations.
The sign of this splitting will depend on the sign of $C_{V1}$,
where  we will have
\begin{eqnarray}
  && M(J=\tfrac{1}{2}) < M(J=\tfrac{3}{2}) \quad \mbox{for $C_{V1} > 0$,}
  \nonumber \\
  && M(J=\tfrac{1}{2}) > M(J=\tfrac{3}{2}) \quad \mbox{for $C_{V1} < 0$, and}
  \nonumber \\
  && M(J=\tfrac{1}{2}) = M(J=\tfrac{3}{2}) \quad \mbox{for $C_{V1} = 0$.}
  \nonumber \\
\end{eqnarray}
We find examples of these three types of hyperfine splitting
in Tables \ref{tab:pot-baryon-meson} and \ref{tab:pot-baryon-antimeson}.

\begin{table}[!ttt]
    \bgroup
\def\arraystretch{1.25}% 
\begin{tabular}{|ccc|ccc|cc|cc|}
\hline\hline
System & S & I & $\lambda^{\bar 3}$ & $\lambda^{6}$ & $\lambda^{\bar 15}$ & $C_0^V$ & $C_1^V$
& $M_{\rm th}$ & $M_{\rm th}^*$ \\
\hline \hline
$N {D}_s$ & $+1$ & $\frac{1}{2}$ & $0$ & $0$ & $1$ & $0$ & $0$ & $2907.3$ & $3051.1$ \\
\hline \hline
$N {D}$ & $0$ & $0$ & $\frac{3}{4}$ & $0$ & $\frac{1}{4}$ & $-6$ & $-6$ & $2806.2$ & $2947.5$ \\
%& & \\
$\Lambda {D}_s$ & $0$ & $0$ & $\frac{1}{4}$ & $0$ & $\frac{3}{4}$ & $-2$  & $-2$ & $3084.1$ & $3227.9$ \\
% & & \\
\hline \hline
$N {D}$ & $0$ & $1$ & $0$ & $\frac{1}{2}$ & $\frac{1}{2}$ & $-2$ & $+\frac{2}{3}$ & $2806.2$ & $2947.5$ \\
$\Sigma {D}_s$ & $0$ & $1$ & $0$ & $\frac{1}{2}$ & $\frac{1}{2}$ & $-2$ & $+\frac{2}{3}$ & $3161.5$ & $3305.4$ \\
\hline \hline
$\Lambda {D}$ & $-1$ & $\tfrac{1}{2}$ & $\frac{1}{16}$ & $\frac{3}{8}$ & $\frac{9}{16}$ & $-2$ & $0$ & $2983.0$ & $3124.3$ \\
% & & \\
$\Sigma {D}$ & $-1$ & $\tfrac{1}{2}$ & $\frac{9}{16}$ & $\frac{3}{8}$ & $\frac{1}{16}$ & $-6$ & $-4$ & $3060.4$ & $3201.7$ \\
$\Xi {D}_s$ & $-1$ & $\tfrac{1}{2}$ & $\frac{3}{8}$ & $\frac{1}{4}$ & $\frac{3}{8}$ & $-4$ & $-\frac{8}{3}$ & $3286.7$ & $3430.5$ \\
\hline \hline
$\Sigma {D}$ & $-1$ & $\tfrac{3}{2}$ & $0$ & $0$ & $1$ & $0$ & $0$
& $3060.4$ & $3201.7$ \\
\hline \hline
$\Xi {D}$ & $-2$ & $0$ & $0$ & $1$ & $0$ & $-4$ & $+\frac{4}{3}$ & $3185.5$ & $3326.9$ \\
\hline \hline
$\Xi {D}$ & $-2$ & $1$ & $0$ & $0$ & $1$ & $0$ & $0$ & $3185.5$ & $3326.9$ \\
\hline \hline
\end{tabular}
\egroup
\caption{SU(3) decomposition of the light octet baryon and
  charmed meson system, which can be decomposed into
  the $8 \otimes \bar{3} = \bar{3} \oplus 6 \oplus \bar{15}$
  representations.
  ``System'' refers to the two-body system under consideration,
  $\lambda_{R}$ the numerical flavor factor for the $V_{R}$ contribution
  to the potential (where $R = \bar{3}$,$6$ or $\bar{15}$),
  $C_0^V$ and $C_1^V$ the relative strength of the electric- and
  magnetic-type piece of vector meson exchange and
  $M_{th}$, $M_{th}^*$ the threshold (in MeV) for
  the system containing a ground ($D$ or $D_s$) or
  excited state ($D^*$ or $D_s^*$) charmed meson.
}
\label{tab:pot-baryon-meson}
\end{table}

\begin{table}[!ttt]
  \bgroup
\def\arraystretch{1.25}% 
\begin{tabular}{|ccc|ccc|cc|cc|}
\hline\hline
System & S & I & $\lambda^3$ & $\lambda^{\bar 6}$ & $\lambda^{15}$ & $C_0^V$ & $C_1^V$
& $M_{\rm th}$ & $M_{\rm th}^*$ \\
\hline \hline
$N\bar{D}$ & $0$ & $0$ & $0$ & $1$ & $0$ & $0$ & $-4$ & $2806.2$ & $2947.5$ \\
\hline \hline
$N\bar{D}$ & $0$ & $1$ & $0$ & $0$ & $1$ & $+4$ & $+\frac{8}{3}$ & $2806.2$ & $2947.5$ \\
\hline \hline
$N\bar{D}_s$ & $-1$ & $\frac{1}{2}$ & $\frac{3}{8}$ & $\frac{1}{4}$ & $\frac{3}{8}$ & $0$ & $0$ & $2907.3$ & $3051.1$ \\
$\Lambda \bar{D}$ & $-1$ & $\frac{1}{2}$ & $\frac{1}{16}$ & $\frac{3}{8}$ &
$\frac{9}{16}$ & $+2$ & $0$ & $2982.9$ & $3124.3$ \\
$\Sigma \bar{D}$ & $-1$ & $\frac{1}{2}$ & $\frac{9}{16}$ & $\frac{3}{8}$ &
$\frac{1}{16}$ & $-2$ & $-\frac{3}{2}$ & $3060.4$ & $3201.7$ \\
\hline \hline
$\Sigma \bar{D}$ & $-1$ & $\frac{3}{2}$ & $0$ & $0$ &
$1$ & $+4$ & $+\frac{8}{3}$ & $3060.4$ & $3201.7$ \\
\hline \hline
$\Lambda \bar{D}_s$ & $-2$ & $0$ & $\frac{1}{4}$ & $0$ & $\frac{3}{4}$ &
$+2$ & $+2$ & $3084.1$ & $3227.9$ \\
$\Xi \bar{D}$ & $-2$ & $0$ & $\frac{3}{4}$ & $0$ & $\frac{1}{4}$ &
$-2$ & $+\frac{2}{3}$ & $3185.5$ & $3326.9$ \\
\hline \hline
$\Xi \bar{D}$ & $-2$ & $1$ & $0$ & $\frac{1}{2}$ & $\frac{1}{2}$ &
$+2$ & $-\frac{2}{3}$ & $3185.5$ & $3326.9$ \\
$\Sigma \bar{D}_s$ & $-2$ & $1$ & $0$ & $\frac{1}{2}$ & $\frac{1}{2}$ &
$+2$ & $-\frac{2}{3}$ & $3161.5$ & $3305.4$ \\
\hline \hline
$\Xi \bar{D}_s$ & $-3$ & $\frac{1}{2}$ & $0$ & $0$ & $1$ & $+4$ & $+\frac{8}{3}$& $3286.7$ & $3430.5$ \\
\hline \hline
\end{tabular}
\egroup
\caption{
SU(3) decomposition of the light octet baryon and
anticharmed meson system, which can be decomposed into
the $8 \otimes {3} = {3} \oplus \bar{6} \oplus {15}$
representations.
We refer to Table \ref{tab:pot-baryon-meson}
for the conventions used here.
}
\label{tab:pot-baryon-antimeson}
\end{table}

\section{Calibration and quantitative predictions}

For calibrating the proportionality constant of the $C^{\rm sat}$ coupling
we need a {\it reference state}, i.e. a molecular candidate from
which we can calculate the coupling by reproducing its mass.
Two suitable choices are the $\Sigma_c(2800)$ and $\Lambda_c(2940)$ charmed
baryons, which have been proposed to be molecular:
\begin{itemize}
\item[(i)] Of the two states, the $\Lambda_c(2940)$ fits the molecular
  interpretation better and is usually interpreted as a $J^P=\tfrac{3}{2}^-$
  $N D^*$ bound state~\cite{He:2006is,He:2010zq,Zhang:2012jk,Ortega:2012cx,Wang:2020dhf,Sakai:2020psu} (though it should be noticed that its $J^P$ is not
  completely established yet).
\item[(ii)] For the $\Sigma_c(2800)$ its interpretation as a molecular
  state is that of a $J^P = \tfrac{1}{2}^-$ $ND$ bound / virtual
  state or resonance~\cite{Haidenbauer:2010ch,Sakai:2020psu},
  but it is more contested~\cite{Wang:2020dhf,Yan:2022nxp}.
\end{itemize}

First, for the calculation of the binding energies we begin by regularizing
the contact-range potential:
\begin{eqnarray}
  \langle \vec{p}\,' | V_C | \vec{p}\, \rangle =
  C^{\rm sat}_{\rm mol}(\Lambda_H)\,
  f(\frac{p'}{\Lambda_H})\,f(\frac{p}{\Lambda_H}) \, ,
\end{eqnarray}
where $f(x)$ is a regularization function and $\Lambda_H$
the regularization scale.
We choose a Gaussian $f(x) = e^{-x^2}$ and a cutoff
$\Lambda_H = 0.75\,{\rm GeV}$ (i.e. close to the vector meson mass).
This potential is inserted into the bound state equation
\begin{eqnarray}
  1 + 2\,\mu_{\rm mol}\,C^{\rm sat}_{\rm mol}(\Lambda_H)\,
  \int_0^{\infty}\,\frac{q^2 dq}{2 \pi^2}\,
  \frac{f^2(q/\Lambda_H)}{q^2 + \gamma_{\rm mol}^2} \, ,
\end{eqnarray}
that is, the Lippmann-Schwinger equation as particularized for the poles of
the scattering amplitude.
Within the bound state equation, $\mu_{\rm mol}$ is the two-body reduced mass
and $\gamma_{\rm mol}$ the wave number of the bound state, which is related % with
to its binding energy $B_{\rm mol}$ by
$\gamma_{\rm mol} = \sqrt{2 \mu_{\rm mol} B_{\rm mol}}$.
Notice that we define $B_{\rm mol} > 0$ for bound states and that the mass of
the molecular state will be given by $M_{\rm mol} = M_{\rm th} - B_{\rm mol}$,
with $M_{\rm th}$ the two-body threshold.
For the regulator we are using, $f(x) = e^{-x^2}$,
the loop integral is given by
\begin{eqnarray}
  && \int_0^{\infty}\,\frac{q^2 dq}{2 \pi^2}\,
  \frac{f^2(q/\Lambda_H)}{q^2 + \gamma_{\rm mol}^2} = \nonumber \\
  && \quad \frac{1}{8 \pi^2}\,\left[
         \sqrt{2 \pi}\,\Lambda_H -
    2\, e^{2 \gamma_{\rm mol}^2 / \Lambda_H^2}\,\pi \gamma_{\rm mol} \,
    {\rm erfc}\left( \frac{\sqrt{2} \gamma_{\rm mol}}{\Lambda_H} \right)
         \right]  \, , \nonumber \\
\end{eqnarray}
with ${\rm erfc}(x)$ the complementary error function.
Depending on the choice of sign for $\gamma_{\rm mol}$, we will talk
about bound ($\gamma_{\rm mol} > 0$) or virtual ($\gamma_{\rm mol} < 0$) states.

The calibration of $C^{\rm sat}_{\rm mol}$ involves its calculation for the
reference state (for which the mass is known),
i.e. we take ``${\rm mol} = {\rm ref}$''.
For the $\Sigma_c(2800)$ and $\Lambda_c(2940)$ cases, this results in
$C^{\rm sat}_{\rm ref} = -1.76\,{\rm fm}^2$ and $-1.74\,{\rm fm}^2$,
respectively (where we use the couplings of Table \ref{tab:couplings}).
For other molecules we define the ratio
\begin{eqnarray}
  R_{\rm mol} =
  \frac{\mu_{\rm mol} C^{\rm sat}_{\rm mol}}{\mu_{\rm ref} C^{\rm sat}_{\rm ref}} \, ,
\end{eqnarray}
which can be determined from Eq.~(\ref{eq:C-sat}) or its SU(3)-flavor extension.
After this, we find the mass of the molecule by solving
\begin{eqnarray}
  1 + (2\mu_{\rm ref} \, C^{\rm sat}_{\rm ref})\,R_{\rm mol}\,
  \int_0^{\infty}\,\frac{q^2 dq}{2 \pi^2}\,
  \frac{f^2(q/\Lambda_H)}{q^2 + \gamma_{\rm mol}^2} = 0 \, .
  \nonumber \\
\end{eqnarray}
This leads to the spectrum we show in Tables \ref{tab:molecular-baryons} and
\ref{tab:molecular-pentaquarks} for the molecular charmed baryons and
anticharmed pentaquarks, respectively.

For the uncertainties, we will do as follows: the largest source of error
in the saturated couplings is the $\sigma$ meson, the parameters and
nature of which are not particularly well known.
Besides, the RG-improved saturated coupling is most sensitive to
the contribution of the $\sigma$ meson owing its lighter mass
when compared to the vector mesons.
Thus we will vary the scalar meson mass within its RPP window of
$m_{\sigma} = (400-550)\,{\rm MeV}$ as a practical method
to estimate the uncertainties of our model.
In addition to this uncertainty there is of course the uncertainty coming
from the choice of a reference state, which results in two different sets
of predictions depending on whether we use the $\Lambda_c(2940)$ or
$\Sigma_c(2800)$.

Regarding the predictions for the molecular baryons
in Table \ref{tab:molecular-baryons}, we find it
worth commenting the following:
\begin{itemize}
\item[(i)] Predictions derived from the $\Sigma_c(2800)$ are considerably
  more attractive than the ones derived from the $\Lambda_c(2940)$.

\item[(ii)] We find molecular matches of the $\Xi_c(3055)$ ($\Sigma D$),
  $\Xi_c(3123)$ ($\Lambda D^*$) and the $\Omega_c(3185)$ ($\Xi D$)
  and $\Omega_c(3327)$ ($J=\tfrac{3}{2}$ $\Xi D^*$)~\cite{LHCb:2023rtu}.

\item[(iii)] The recent LHCb manuscript in which the $\Omega_c(3185/3327)$
  have been discovered~\cite{LHCb:2023rtu} also indicates that
  no structures have been observed in $\Xi_c^+ K^+$.
  $\Sigma D_s^{(*)}$ molecules can decay into this channel via a short-range
  operator (exchange of a light-baryon). Though only expected to generate
  a narrow width, the size of this matrix element grows
  with the binding energy~\footnote{More binding implies a larger probability
    of the two particles being close to each other, which for a short-range
    operator would be a necessary  condition for having
    a non-negligible matrix element.},
  disfavoring the use of  $\Sigma_c(2800)$ as a reference state
  because of the large bindings it entails for $\Sigma D_s^{(*)}$.

\item[(iv)] Curiously, if $\Sigma_c(2800)$ is the reference state, we predict
  two $I=0$ $N D^*$ bound states that might correspond to the
  $\Lambda_c(2940)$ (but now appearing as a $J=\tfrac{1}{2}$ state) and
  the recently discovered $\Lambda_c(2910)$~\cite{Belle:2022hnm}
  (as a $J=\tfrac{3}{2}$ state).
  This interpretation coincides with the one proposed in~\cite{Zhang:2022pxc},
  but not with Refs.~\cite{Azizi:2022dpn,Yan:2022nxp} that consider
  the $\Lambda_c(2910)$ as compact or at least non-molecular.
\end{itemize}

If we consider the anticharmed pentaquarks predicted
in Table~\ref{tab:molecular-pentaquarks},
the first problem we are confronted with is the lack of candidates.
Nonetheless, there is experimental information about $I=0$ $N \bar{D}$
scattering at low energies from the ALICE collaboration~\cite{ALICE:2022enj},
which constrained the values of the inverse scattering
length~\footnote{In~\cite{ALICE:2022enj} the sign convention of the scattering
  length is $f_0 > 0$ ($f_0 < 0$) for a two-body system with a virtual
  (bound) state.} of this system to the following range:
\begin{eqnarray}
  f_0^{-1}(I=0) \in [-0.4,0.9] \, {\rm fm}^{-1} \, .
\end{eqnarray}
The calculation of $f_0$ in our formalism is given by
\begin{eqnarray}
  -\frac{1}{f_0} =
  \frac{2\pi}{\mu_{\rm ref}\,C^{\rm sat}_{\rm ref}}\,\frac{1}{R_{\rm mol}}
  + \frac{2}{\pi}\,\int_0^{\infty} \, dq \, f^2(\frac{q}{\Lambda_H}) \, ,
\end{eqnarray}
and, depending on the reference state used, we arrive at
\begin{eqnarray}
  f_0^{-1}(I=0) &=& +0.34\, {\rm fm}^{-1} \quad \mbox{for ref = $\Lambda_c(2940)$} \, , \nonumber \\
  f_0^{-1}(I=0) &=& -0.24\, {\rm fm}^{-1} \quad \mbox{for ref = $\Sigma_c(2800)$} \, . \nonumber \\ 
\end{eqnarray}
That is, from the prediction of the inverse scattering length
we conclude that both reference states comply
with this experimental constraint.

Alternatively, we might compare the spectrum
in Table~\ref{tab:molecular-pentaquarks}
with previous theoretical predictions.
The first predictions of a $\bar{c} qqqq$ pentaquark are maybe the ones by
Gignoux et al.~\cite{Gignoux:1987cn} and Lipkin~\cite{Lipkin:1987sk},
who calculated that the anticharmed-strange pentaquark configurations
could be stable and located below the $N \bar{D}_s$ threshold.
Here the $N \bar{D}_s$ system shows a remarkable amount of attraction,
but binding is subordinate to our choice of reference state:
from the $\Sigma_c(2800)$ we indeed find a shallow bound state,
but if we use the $\Lambda_c(2940)$ instead, we end up with
a virtual state (albeit close to threshold).
Hofmann and Lutz~\cite{Hofmann:2005sw} proposed that
the $N \bar{D}_s$-$\Lambda \bar{D}$-$\Sigma \bar{D}$ and
$\Lambda \bar{D}_s$-$\Xi \bar{D}$ systems might generate bound states at
$2.78$ and $2.84\,{\rm GeV}$, respectively (and also a hidden-charmed
pentaquark at $3.52\,{\rm GeV}$, probably one of the first
predictions of these states).
Even though we find considerably less attraction for the aforementioned
systems than in~\cite{Hofmann:2005sw}, these systems are still
attractive and able to bind within our model.
More recently, Yalikun and Zou~\cite{Yalikun:2021dpk} have studied
possible $\Sigma \bar{D}$ and $\Sigma \bar{D}^*$ bound states
within the one boson exchange model.
We find three possible near-threshold states in these configurations
in agreement with~\cite{Yalikun:2021dpk}.
That is, in general the qualitative features of the spectrum we predict align
with previous results, though there are differences at the quantitative level,
which will only be dilucidated once we have further experimental results.

\begin{table*}[!ttt]
\begin{tabular}{|c||ccc||ccc||ccc||cc|}
\hline\hline
System & $S$ & $I$ & $J^P$ &
$R_{\rm mol}$(${\Lambda_c^*}$) & $B_{\rm mol}$ & $M_{\rm mol}$ &
$R_{\rm mol}$(${\Sigma_c^*}$) & $B_{\rm mol}$ & $M_{\rm mol}$ &
Candidate & $M_{\rm cand}$
\\
\hline \hline
$N D_s$ & $+1$ & $\frac{1}{2}$ & $\frac{1}{2}^-$
& $0.60$ & $(2.2)^V$ & $(2905.0^{+2.2(B)}_{-6.9})^V$
& $0.91$ & $2.8$ & $2904.5 \pm 1.4$ & - & - \\
% \hline
$N D_s^*$ & $+1$ & $\frac{1}{2}$ & $\frac{1}{2}^-$
& $0.62$ & $(1.7)^V$ & $(3049.4^{+1.7(B)}_{-6.3})^V$
& $0.95$ & $3.4$ & $3047.7^{+1.6}_{-1.5}$ & - & - \\
$N D_s^*$ & $+1$ & $\frac{1}{2}$ & $\frac{3}{2}^-$
& $0.62$ & $(1.7)^V$ & $(3049.4^{+1.7(B)}_{-6.4})^V$
& $0.95$ & $3.4$ & $3047.7^{+1.6}_{-1.5}$ & - & - \\
\hline
\hline
$N D$ & $0$ & $0$ & $\frac{1}{2}^-$
& $0.79$ & $0.6$ & $2805.6^{+0.5}_{-1.2}$ 
& $1.20$ & $17.7$ & $2788.5^{+4.9}_{-6.8}$
& $\Lambda_c(2765)$ & $2766.6 \pm 2.4$~\cite{ParticleDataGroup:2022pth} \\
% \hline
$N D^*$ & $0$ & $0$ & $\frac{1}{2}^-$
& $0.44$ & $(16)^V$ & ${(2932^{+14}_{-34})}^V$
% & $0.44$ & $(15.9)^V$ & ${(2931.6^{+13.5}_{-33.9})}^V$
& $0.66$ & $1.2$ & $2946.3^{+1.0}_{-10.5}$ &
$\Lambda_c(2940)$ & $2939.6^{+1.3}_{-1.5}$~\cite{ParticleDataGroup:2022pth} \\
$N D^*$ & $0$ & $0$ & $\frac{3}{2}^-$ &
$1$ (Input) & $7.9$ & $2939.6$
% & $1.51$ & $41.7$ & $2905.8^{+15.3}_{-21.6}$
& $1.51$ & $42$ & $2906^{+15}_{-22}$
& $\Lambda_c(2940)$ & $2939.6^{+1.3}_{-1.5}$~\cite{ParticleDataGroup:2022pth} \\
\hline \hline
$N D$ & $0$ & $1$ & $\frac{1}{2}^-$
& $0.66$ & $(0.6)^V$ & $(2805.6^{+0.6(B)}_{-3.1})^V$
& $1$ (Input) & $6.2$ & $2800.0$ & $\Sigma_c(2800)$ & $\sim 2800$~\cite{ParticleDataGroup:2022pth} \\
% \hline
$N D^*$ & $0$ & $1$ & $\frac{1}{2}^-$
& $0.72$ & $(0.0)^V$ & $(2947.5^{+0.0(B)}_{-1.1})^V$
& $1.09$  & $10.4$ & $2937.1^{+1.3}_{-1.7}$ & & \\
$N D^*$ & $0$ & $1$ & $\frac{3}{2}^-$
& $0.66$ & $(0.7)^V$ & $(2946.8^{+0.7(B)}_{-3.7})^V$
& $0.99$ & $5.7$ & $2941.8^{+0.6}_{-0.5}$ & & \\
\hline \hline
$\Lambda D_s$ & $0$ & $0$ & $\frac{1}{2}^-$
& $0.54$ & $(5.0)^V$ & $(3079.0^{+3.7}_{-1.9})^V$
& $0.82$ & $0.4$ & $3083.6 \pm 0.2$ & & \\
% \hline
$\Lambda D_s^*$ & $0$ & $0$ & $\frac{1}{2}^-$
& $0.51$ & $(7.0)^V$ & $(3220.9^{+5.2}_{-2.1})^V$
& $0.77$ & $0.0$ & $3227.9^{+0.0(V)}_{-0.3}$ & & \\
$\Lambda D_s^*$ & $0$ & $0$ & $\frac{3}{2}^-$
& $0.59$ & $(3.0)^V$ & $(3224.9^{+2.5}_{-2.4})^V$
& $0.87$ & $1.4$ & $3226.5^{+0.0(V)}_{-0.3}$ & & \\
\hline \hline
$\Sigma D_s$ & $0$ & $1$ & $\frac{1}{2}^-$
& $0.74$ & $0.0$ & $3164.5^{+0.0(V)}_{-1.9}$
& $1.12$ & $10.6$ & $3150.9^{+1.6}_{-1.3}$ & - & - \\
% \hline
$\Sigma D_s^*$ & $0$ & $1$ & $\frac{1}{2}^-$
& $0.74$ & $0.0$ & $3305.3^{+0.0(V)}_{-2.1}$
& $1.13$ & $10.7$ & $3294.7^{+2.1}_{-1.8}$ & - & - \\
$\Sigma D_s^*$ & $0$ & $1$ & $\frac{3}{2}^-$
& $0.77$ & $0.2$ & $3305.2^{+0.2(V)}_{-2.4}$
& $1.16$ & $12.5$ & $3292.9^{+1.4}_{-1.2}$ & - & - \\
\hline \hline
$\Lambda D$ & $-1$ & $\frac{1}{2}$ & $\frac{1}{2}^-$
& $0.57$ & $(3.4)^V$ & $(2979.6^{+2.5}_{-4.3})^V$
& $0.87$ & $1.3$ & $2981.7^{+0.2}_{-0.3}$ & - & - \\
% \hline
$\Lambda D^*$ & $-1$ & $\frac{1}{2}$ & $\frac{1}{2}^-$
& $0.59$ & $(2.6)^V$ & $(3121.6^{+2.1}_{-3.9})^V$
& $0.89$ & $1.8$ & $3122.5^{+0.3}_{-0.4}$ & $\Xi_c(3123)$ & $3122.9 \pm 1.3$~\cite{ParticleDataGroup:2022pth}\\
$\Lambda D^*$ & $-1$ & $\frac{1}{2}$ & $\frac{3}{2}^-$
& $0.59$ & $(2.6)^V$ & $(3121.6^{+2.1}_{-3.9})^V$
& $0.89$ & $1.8$ & $3122.5^{+0.3}_{-0.4}$ & $\Xi_c(3123)$ & $3122.9 \pm 1.3$~\cite{ParticleDataGroup:2022pth} \\
\hline
\hline
$\Sigma D$ & $-1$ & $\frac{1}{2}$ & $\frac{1}{2}^-$
& $0.92$ & $3.8$ & $3056.6^{+1.9}_{-2.5}$
& $1.40$ & $28.1$ & $3023.3^{+6.1}_{-8.2}$ &
$\Xi_c(3055)$ & $3055.9 \pm 0.4$~\cite{ParticleDataGroup:2022pth} \\
% \hline
$\Sigma D^*$ & $-1$ & $\frac{1}{2}$ & $\frac{1}{2}^-$
& $0.66$ & $(0.6)^V$ & $(3201.1^{+0.6(B)}_{-5.8})^V$
& $0.99$ & $5.0$ & $3196.8^{+3.0}_{-3.2}$ & - & - \\
$\Sigma D^*$ & $-1$ & $\frac{1}{2}$ & $\frac{3}{2}^-$
& $1.10$ & $11.5$ & $3190.3^{+1.1}_{-1.2}$
& $1.66$ & $47$ & $3155^{+14}_{-18}$ & - & -\\
% & $1.66$ & $47.0$ & $3154.7^{+13.5}_{-18.4}$ & - & -\\
\hline \hline
$\Sigma D$ & $-1$ & $\frac{3}{2}$ & $\frac{1}{2}^-$
& $0.69$ & $(0.1)^V$ & $(3060.3^{+0.1(B)}_{-3.4})^V$
& $1.05$ & $7.3$ & $3053.1^{+2.3}_{-2.1}$   & - & - \\
% \hline
$\Sigma D^*$ & $-1$ & $\frac{3}{2}$ & $\frac{1}{2}^-$
& $0.71$ & $(0.0)^V$ & $(3201.7^{+0.0(B)}_{-2.7})^V$
& $1.08$ & $8.4$ & $3193.3^{+2.5}_{-2.2}$ & - & - \\
$\Sigma D^*$ & $-1$ & $\frac{3}{2}$ & $\frac{1}{2}^-$
& $0.71$ & $(0.0)^V$ & $(3201.7^{+0.0(B)}_{-2.7})^V$
& $1.08$ & $8.4$ & $3193.3^{+2.5}_{-2.2}$ & - & - \\
\hline \hline
$\Xi D_s$ & $-1$ & $\frac{1}{2}$ & $\frac{1}{2}^-$
& $0.82$ & $1.0$ & $3286.7^{+0.9}_{-3.2}$
& $1.25$ & $16.6$ & $3270.0^{+0.4}_{-0.3}$ & - & - \\
% \hline
$\Xi D_s^*$ & $-1$ & $\frac{1}{2}$ & $\frac{1}{2}^-$
& $0.91$ & $3.1$ & $3427.4^{+2.3}_{-3.8}$
& $1.38$ & $24.2$ & $3406.3^{+2.0}_{-2.5}$ & - & - \\
$\Xi D_s^*$ & $-1$ & $\frac{1}{2}$ & $\frac{3}{2}^-$
& $0.81$ & $0.8$ & $3429.8^{+0.6}_{-3.4}$
& $1.23$ & $15.3$ & $3415.2^{+1.6}_{-1.4}$ & - & - \\
\hline \hline
$\Xi D$ & $-2$ & $0$ & $\frac{1}{2}^-$
& $0.90$ & $2.8$ & $3182.7^{+2.1}_{-3.2}$
& $1.36$ & $23.9$ & $3161.6^{+3.0}_{-3.9}$
& $\Omega_c(3185)$ & $3185.1^{+7.6}_{-1.9}$~\cite{LHCb:2023rtu} \\
% \hline
$\Xi D^*$ & $-2$ & $0$ & $\frac{1}{2}^-$
& $1.03$ & $7.6$ & $3319.3^{+2.5}_{-3.0}$
& $1.56$ & $36.7$ & $3290.2^{+7.5}_{-10.0}$ & - & - \\
$\Xi D^*$ & $-2$ & $0$ & $\frac{3}{2}^-$
& $0.87$ & $2.0$ & $3324.8^{+1.9}_{-3.6}$
& $1.32$ & $20.9$ & $3306.0^{+1.1}_{-1.4}$
& $\Omega_c(3327)$ & $3327.1^{+1.2}_{-1.8}$~\cite{LHCb:2023rtu} \\
\hline \hline
$\Xi D$ & $-2$ & $1$ & $\frac{1}{2}^-$
& $0.73$ & $0.0$ & $3185.5^{+0.0(V)}_{-2.1}$
& $1.11$ & $9.8$ & $3175.8^{+2.6}_{-2.3}$ & - & - \\
% \hline
$\Xi D^*$ & $-2$ & $1$ & $\frac{1}{2}^-$
& $0.76$ & $0.1$ & $3326.8^{+0.1(V)}_{-2.6}$
& $1.15$ & $11.1$ & $3315.8^{+2.8}_{-2.5}$ & - & -\\
$\Xi D^*$ & $-2$ & $1$ & $\frac{1}{2}^-$
& $0.76$ & $0.1$ & $3326.8^{+0.1(V)}_{-2.6}$
& $1.15$ & $11.1$ & $3315.8^{+2.8}_{-2.5}$ & - & - \\
\hline \hline
\end{tabular}
\caption{Molecular charmed baryons predicted in our model.
  ``System'' refers to the octet baryon - charmed meson pair under
  consideration, $S$, $I$, $J^P$ to their strangeness, isospin and spin-parity,
  $R_{\rm mol}$ to the relative strength (central value) of
  the saturated coupling with respect to the $\Lambda_c(2940)$ or
  $\Sigma_c(2800)$ as $N D^{(*)}$ molecules,
  $B_{\rm mol}$ to the binding energy (central value), $M_{\rm mol}$ to the
  mass of the molecule (includes uncertainties), ``Candidate'' to a possible
  molecular candidate corresponding to the configuration we are calculating,
  and $M_{\rm cand}$ to the mass of this candidate.
  A superscript $V$ over the binding energy or mass indicates a virtual
  state solution.
  The uncertainties in $M_{\rm mol}$ come from varying the scalar meson mass
  in the $(400-550)\,{\rm MeV}$ range (while a change in the sheet, e.g.
  from virtual to bound, is indicated with a $B$ or $V$ superscript
  in parentheses and next to the error).
  All binding energies and masses are in units of ${\rm MeV}$.
}
\label{tab:molecular-baryons}
\end{table*}

\begin{table*}[!ttt]
\begin{tabular}{|c||ccc||ccc||ccc|}
\hline\hline
System & $S$ & $I$ & $J^P$ &
$R_{\rm mol}$(${\Lambda_c^*}$) & $B_{\rm mol}$ & $M_{\rm mol}$ &
$R_{\rm mol}$(${\Sigma_c^*}$) & $B_{\rm mol}$ & $M_{\rm mol}$
\\
\hline
\hline
$N \bar{D}$ & $0$ & $0$ & $\frac{1}{2}^-$
& $0.59$ & $(2.7)^V$ & $(2803.4^{+2.6}_{-7.4})^V$
& $0.90$ & $2.3$ & $2803.8^{+1.2}_{-1.3}$ \\
%\hline
$N \bar{D}^*$ & $0$ & $0$ & $\frac{1}{2}^-$
& $0.36$ & $(30)^V$ & $(2918^{+24}_{-68})^V$
% & $0.36$ & $(29.9)^V$ & $(2917.6^{+23.9}_{-67.9})^V$
& $0.54$ & $(7.0)^V$ & $2940.5^{+6.9}_{-35.9}$ \\
$N \bar{D}^*$ & $0$ & $0$ & $\frac{3}{2}^-$
& $0.73$ & $0.0$ & $2947.5^{+0.0(V)}_{-0.9}$
& $1.11$ & $11.7$ & $2935.8^{+1.8}_{-2.4}$ \\
\hline \hline
$N \bar{D}$ & $0$ & $1$ & $\frac{1}{2}^-$
& $0.46$ & $(13)^V$ & $(2793^{+11}_{-26})^V$
% & $0.46$ & $(13.1)^V$ & $(2793.1^{+11.2}_{-26.2})^V$
& $0.70$ & $(0.4)^V$ & $(2805.8_{-5.7}^{+0.4(B)})^V$ \\
%\hline
$N \bar{D}^*$ & $0$ & $1$ & $\frac{1}{2}^-$
& $0.64$ & $(1.1)^V$ & $(2946.4^{+1.1(B)}_{-4.6})^V$
& $0.97$ & $4.7$ & $2942.8^{+1.0}_{-0.9}$ \\
$N \bar{D}^*$ & $0$ & $1$ & $\frac{3}{2}^-$
& $0.39$ & $(23)^V$ & $(2924^{+19}_{-51})^V$
% & $0.39$ & $(23.2)^V$ & $(2924.3^{+19.2}_{-51.2})^V$
& $0.59$ & $(4.0)^V$ & $(2943.4^{+4.0(B)}_{-22.7})^V$ \\
\hline \hline
$N \bar{D}_s$ & $-1$ & $\frac{1}{2}$ & $\frac{1}{2}^-$
& $0.60$ & $(2.2)^V$ & $(2905.0^{+2.2(B)}_{-6.9})^V$
& $0.92$ & $2.7$ & $2904.5 \pm 1.4$ \\
% \hline
$N \bar{D}_s^*$ & $-1$ & $\frac{1}{2}$ & $\frac{1}{2}^-$
& $0.62$ & $(1.7)^V$ & $(3049.4^{+1.7(B)}_{-6.3})^V$
& $0.94$ & $3.4$ & $3047.7 \pm 1.5$ \\
$N \bar{D}_s^*$ & $-1$ & $\frac{1}{2}$ & $\frac{3}{2}^-$
& $0.62$ & $(1.7)^V$ & $(3049.4^{+1.7(B)}_{-6.3})^V$
& $0.94$ & $3.4$ & $3047.7 \pm 1.5$ \\
\hline \hline
$\Lambda \bar{D}$ & $-1$ & $\frac{1}{2}$ & $\frac{1}{2}^-$
& $0.42$ & $(16)^V$ & $(2967^{+11}_{-20})^V$
% & $0.42$ & $(16.1)^V$ & $(2966.9^{+10.9}_{-20.1})^V$
& $0.64$ & $(1.6)^V$ & $(2981.4^{+1.4}_{-4.3})^V$ \\
%\hline
$\Lambda \bar{D}^*$ & $-1$ & $\frac{1}{2}$ & $\frac{1}{2}^-$
& $0.44$ & $(14)^V$ & $(3110^{+10}_{-19})^V$
% & $0.44$ & $(14.3)^V$ & $(3110.0^{+10.0}_{-19.0})^V$
& $0.66$ & $(1.1)^V$ & $(3123.2^{+1.1(B)}_{-3.8})^V$ \\
$\Lambda \bar{D}^*$ & $-1$ & $\frac{1}{2}$ & $\frac{3}{2}^-$
& $0.44$ & $(14)^V$ & $(3110^{+10}_{-19})^V$
% & $0.44$ & $(14.3)^V$ & $(3110.0^{+10.0}_{-19.0})^V$
& $0.66$ & $(1.1)^V$ & $(3123.2^{+1.1(B)}_{-3.8})^V$ \\
\hline
\hline
$\Sigma \bar{D}$ & $-1$ & $\frac{1}{2}$ & $\frac{1}{2}^-$
& $0.77$ & $0.2$ & $3060.2^{+0.2(V)}_{-2.1}$
& $1.17$ & $13.2$ & $3047.2 \pm 0$ \\
%\hline
$\Sigma \bar{D}^*$ & $-1$ & $\frac{1}{2}$ & $\frac{1}{2}^-$
& $0.69$ & $(0.1)^V$ & $(3201.6^{+0.1(B)}_{-3.6})^V$
& $1.05$ & $7.2$ & $3194.5^{+2.8}_{-2.6}$ \\
$\Sigma \bar{D}^*$ & $-1$ & $\frac{1}{2}$ & $\frac{3}{2}^-$
& $0.84$ & $1.3$ & $3200.4^{+1.3(V)}_{-2.9}$
& $1.27$ & $19.0$ & $3182.7^{+1.7}_{-2.2}$ \\
\hline \hline
$\Sigma \bar{D}$ & $-1$ & $\frac{3}{2}$ & $\frac{1}{2}^-$
& $0.54$ & $(5.1)^V$ & ${(3055.3^{+5.1(B)}_{-18.1})}^V$
& $0.81$ & $0.4$ & $3060.0^{+0.4(V)}_{-3.0}$ \\
%\hline
$\Sigma \bar{D}^*$ & $-1$ & $\frac{3}{2}$ & $\frac{1}{2}^-$
& $0.75$ & $0.1$ & $3201.7^{+0.1(B)}_{-2.0}$
& $1.13$ & $11.1$ & $3190.6^{+1.5}_{-1.3}$ \\
$\Sigma \bar{D}^*$ & $-1$ & $\frac{3}{2}$ & $\frac{1}{2}^-$
& $0.45$ & $(12)^V$ & ${(3190^{+11}_{-38})}^V$
% & $0.45$ & $(11.5)^V$ & ${(3190.2^{+11.1}_{-38.1})}^V$
& $0.69$ & $(0.5)^V$ & $(3201.2^{+0.5(B)}_{-13.6})^V$ \\
\hline \hline
$\Lambda \bar{D}_s$ & $-2$ & $0$ & $\frac{1}{2}^-$
& $0.47$ & $(10.1)^V$ & $(3073.9^{+7.1}_{-13.0})^V$
& $0.72$ & $(0.1)^V$ & $(3083.9^{+0.1(B)}_{-1.1})^V$ \\
% \hline
$\Lambda \bar{D}_s^*$ & $-2$ & $0$ & $\frac{1}{2}^-$
& $0.53$ & $(5.5)^V$ & $(3222.4^{+4.2}_{-7.8})^V$
& $0.80$ & $0.3$ & $3227.6^{+0.2}_{-0.3}$ \\
$\Lambda \bar{D}_s^*$ & $-2$ & $0$ & $\frac{3}{2}^-$
& $0.46$ & $(10.9)^V$ & $(3217.0^{+7.9}_{-15.0})^V$
& $0.70$ & $(0.3)^V$ & $(3227.6^{+0.3(B)}_{-1.9})^V$ \\
\hline \hline
$\Sigma \bar{D}_s$ & $-2$ & $1$ & $\frac{1}{2}^-$
& $0.67$ & $(0.4)^V$ & $(3161.2^{+0.4(B)}_{-4.7})^V$
& $1.02$ & $5.9$ & $3155.6^{+3.6}_{-2.8}$ \\
% \hline
$\Sigma \bar{D}_s^*$ & $-2$ & $1$ & $\frac{1}{2}^-$
& $0.70$ & $(0.0)^V$ & $(3305.3^{+0.0(B)}_{-3.1})^V$
& $1.07$ & $7.8$ & $3297.5^{+3.1}_{-2.7}$ \\
$\Sigma \bar{D}_s^*$ & $-2$ & $1$ & $\frac{3}{2}^-$
& $0.68$ & $(0.2)^V$ & $(3305.1^{+0.2(B)}_{-4.5})^V$
& $1.03$ & $6.3$ & $3299.0^{+3.2}_{-3.1}$ \\
\hline \hline
$\Xi \bar{D}$ & $-2$ & $0$ & $\frac{1}{2}^-$
& $0.82$ & $0.8$ & $3184.2^{+0.8(V)}_{-3.0}$
& $1.24$ & $16.3$ & $3169.2 \pm 0$ \\
% \hline
$\Xi \bar{D}^*$ & $-2$ & $0$ & $\frac{1}{2}^-$
& $0.89$ & $2.6$ & $3324.3^{+2.1}_{-3.6}$
& $1.35$ & $22.7$ & $3304.1^{+1.8}_{-2.4}$ \\
$\Xi \bar{D}^*$ & $-2$ & $0$ & $\frac{3}{2}^-$
& $0.82$ & $0.8$ & $3326.1^{+0.6}_{-3.3}$
& $1.23$ & $15.7$ & $3311.1^{+1.0}_{-0.8}$ \\
\hline \hline
$\Xi \bar{D}$ & $-2$ & $1$ & $\frac{1}{2}^-$
& $0.65$ & $(0.6)^V$ & $(3184.9^{+0.6(B)}_{-6.7})^V$
& $0.99$ & $4.6$ & $3181.0^{+3.6}_{-3.9}$ \\
%\hline
$\Xi \bar{D}^*$ & $-2$ & $1$ & $\frac{1}{2}^-$
& $0.62$ & $(1.3)^V$ & $(3325.5^{+0.8}_{-10.2})^V$
& $0.94$ & $2.9$ & $3324.0^{+3.1}_{-4.8}$ \\
$\Xi \bar{D}^*$ & $-2$ & $1$ & $\frac{1}{2}^-$
& $0.70$ & $(0.1)^V$ & $(3326.8^{+0.1(B)}_{-4.1})^V$
& $1.06$ & $7.1$ & $3319.8^{+3.7}_{-3.5}$ \\
\hline \hline
$\Xi \bar{D}_s$ & $-3$ & $\frac{1}{2}$ & $\frac{1}{2}^-$
& $0.68$ & $(0.2)^V$ & $(3286.4^{+0.2(B)}_{-5.3})^V$
& $1.03$ & $6.0$ & $3280.7^{+3.3}_{-3.8}$ \\
%\hline
$\Xi \bar{D}_s^*$ & $-3$ & $\frac{1}{2}$ & $\frac{1}{2}^-$
& $0.76$ & $0.1$ & $3430.4_{-9.7}^{+0.1(B)}$
& $1.15$ & $11.5$ & $3419.0^{+2.8}_{-4.5}$ \\
$\Xi \bar{D}_s^*$ & $-3$ & $\frac{1}{2}$ & $\frac{3}{2}^-$
& $0.66$ & $(0.4)^V$ & $(3430.1^{+0.4(B)}_{-2.6})$
& $1.00$ & $5.0$ & $3425.5 \pm 3.7$ \\
\hline \hline
\end{tabular}
\caption{Molecular anticharmed pentaquarks predicted in our model.
  We refer to Table \ref{tab:molecular-baryons} for the conventions used,
  where the only significant difference with the aforementioned Table
  is that here there are no experimental candidates (and hence
  we do not include the ``Candidate'' and $M_{\rm cand}$ columns).
  All binding energies and masses are in units of ${\rm MeV}$.
}
\label{tab:molecular-pentaquarks}
\end{table*}

\section{Isospin breaking effects and the $\Omega_c(3185/3327)$}

The previous predictions have been done in the isospin symmetric limit,
i.e. our calculations use the isospin averages of the charmed meson
and light octet baryon masses.

The inclusion of explicit isospin breaking effects will have different effects
depending on the particular two-body system under consideration.
The effects are trivial in meson-baryon systems for which there is only
one particle channel per isospin state (e.g. $\Xi D_s$ for which
the two isospin states are
$|\frac{1}{2} \frac{1}{2} \rangle_I = | \Xi^0 D_s^+ \rangle$ and
$|\frac{1}{2} -\frac{1}{2} \rangle_I = | \Xi^- D_s^+ \rangle$).
Here isospin breaking only entails a shift in the mass of the molecule
equal to the shift of the physical and isospin symmetric thresholds
(e.g. $\pm 3.4\,{\rm MeV}$ for the $\Xi^- D_s^+$ and $\Xi^0 D_s^+$
molecules, with respect to the $\Xi D_s$ calculations of
Table ~\ref{tab:molecular-baryons}).

More interesting is the case of the $N D$ and $\Xi D$ systems, for which
isospin mixing of the $I=0$ and $I=1$ states is possible (or
the $\Sigma D$ system, where mixing happens between the
$I=\tfrac{1}{2}$ and $I = \tfrac{3}{2}$ configurations,
though we will not consider this case in detail here).
For $N D$ and $\Xi D$ with $M_I=0$ (with $M_I$ the third component of
the isospin wave function)
we have a {\it light} and {\it heavy} particle channel
\begin{eqnarray}
  | 0 0 \rangle_I =
  \frac{1}{\sqrt{2}}\,\left[ | L \rangle - |H \rangle \right] \, ,
  \label{eq:isospin-0} \\
  | 1 0 \rangle_I =
  \frac{1}{\sqrt{2}}\,\left[ | L \rangle + |H \rangle \right] \, ,
  \label{eq:isospin-1}
\end{eqnarray}
where $| L \rangle = | p D^0 \rangle$ or $| \Xi^0 D^0 \rangle$ and
$| H \rangle = - | n D^+ \rangle$ or $- | \Xi^- D^+ \rangle$~\footnote{Here
  we are making use of the existence of a relative sign
  for the isospin states of the light antiquarks:
  $| \bar{d} \rangle = - | \tfrac{1}{2} \tfrac{1}{2} \rangle_I$ and
  $| \bar{u} \rangle = | \tfrac{1}{2} -\tfrac{1}{2} \rangle_I$.
  If we extend this convention to the charmed mesons, which contain
  an antiquark, we arrive at the minus sign for the definition of
  the $| H \rangle$ state.
},
depending on the system.
This decomposition implies that the contact-range potential now
becomes a matrix in the $\{ |L\rangle, |H\rangle \}$ basis.
The identity and product isospin operators change to
\begin{eqnarray}
  1 \to \begin{pmatrix} +1 & 0 \\ 0 & +1 \end{pmatrix}
  \quad \mbox{and} \quad \hat{T}_{12} \to
  \begin{pmatrix} +1 & -2 \\ -2 & +1 \end{pmatrix}
  \, ,
\end{eqnarray}
from which the explicit expression of the saturated
contact-range potential reads
\begin{eqnarray}
  && C^{\rm sat}(m_V) \propto \nonumber \\
  && \quad
  \begin{pmatrix} +1 & -2 \\ -2 & +1 \end{pmatrix}\,
  \frac{g_{\rho 1} g_{\rho 2}}{m_V^2}\,\left( 1 + \kappa_{\rho 1} \kappa_{\rho 2}\,\frac{m_V^2}{6 M^2}\,\hat{S}_{L12} \right) + \nonumber \\ && \quad
  \begin{pmatrix} +\zeta & 0 \\ 0 & +\zeta \end{pmatrix}\,
  \frac{g_{\omega 1} g_{\omega 2}}{m_{V}^2}\,\left( 1 + \kappa_{\omega 1} \kappa_{\omega 2}\,\frac{m_V^2}{6 M^2}\,\hat{S}_{L12} \right) + \nonumber \\ && \quad
  \begin{pmatrix} -1 & 0 \\ 0 & -1 \end{pmatrix}\,
  \left( \frac{m_V}{m_S}\right)\,\frac{g_{S1} g_{S2}}{m_S^2} \, ,
\end{eqnarray}
where it is apparent that the isospin breaking effects derive
from $\rho$ exchange between the $L$ and $H$ channels.

The bound state equation becomes now a two-channel linear system
\begin{eqnarray}
  \phi_A + 2\mu_{\rm ref} \sum_{B} \phi_B\,C^{\rm sat}_{\rm ref}(\Lambda)\,R_{\rm mol}^{AB}\,\int \frac{q^2 d q}{2 \pi^2}\,
  \frac{f^2(\frac{q}{\Lambda})}{q^2 + \gamma_{{\rm mol}(A)}^2}
   = 0\, , \nonumber \\
\end{eqnarray}
where $A,B = L,H$ are indices denoting the channels,
$\phi_A$ the vertex function for channel $A$,
$\gamma_{{\rm mol}(A)} = \sqrt{2 \mu_A (M_{{\rm th}(A)} - M_{\rm mol})}$ with
$M_{\rm mol}$ the mass of the predicted molecule, $M_{{\rm th}(A)}$ the
mass of threshold $A$ and $\mu_A$ the reduced mass of
channel $A$.
The ratio $R_{\rm mol}^{AB}$ is given by
\begin{eqnarray}
  R_{\rm mol}^{AB} =
  \frac{\mu_A C^{\rm sat}_{{\rm mol}(AB)}(\Lambda)}
       {\mu_{\rm ref} C^{\rm sat}_{\rm ref}(\Lambda)} \, ,
       \label{eq:Rmol-AB}
\end{eqnarray}
where the indices $AB$ in the saturated coupling refer to the components of
$C^{\rm sat}$ in matrix form for a given molecule ``mol''.
For simplicity, $C^{\rm sat}_{\rm ref}$ will refer to the coupling of
the reference state in the isospin symmetric limit.
For the $M_I = 0$ $ND$ and $\Xi D$ systems --- i.e. the states with
third component of their isospin wave function
$| I M_I \rangle$ equal zero, check Eqs. (\ref{eq:isospin-0})
and (\ref{eq:isospin-1}) --- the $I=0$ and $I=1$ configurations
correspond to the vertex functions
\begin{eqnarray}
  \phi(I=0) = ( \phi_L, \phi_H ) &=&
  ( \frac{1}{\sqrt{2}}, -\frac{1}{\sqrt{2}} ) \, , \\
  \phi(I=1) = ( \phi_L, \phi_H ) &=&
  ( \frac{1}{\sqrt{2}}, +\frac{1}{\sqrt{2}} ) \, .
\end{eqnarray}

Owing to the different masses of the $L$ and $H$ channels,
the $I=0$ and $I=1$ configurations will mix.
In turn, this will entail changes in the predicted masses.
Naively, the size of this effect is expected to be of the order of the ratio
of the binding energy over the mass gap of the $L$ and
$H$ channels.
However, in practice what we find is that if in the isospin symmetric limit
the molecular state is predicted below the threshold of the $L$ channel,
the impact of isospin breaking in its mass will be rather small.

With the previous formalism we can estimate the effects of isospin
breaking in the two reference states :
\begin{itemize}
\item[(i)] We first calculate $C^{\rm sat}_{\rm ref}$ in the isospin limit
  for a given reference state.
\item[(ii)] Then we recalculate the mass of said reference state after
  the inclusion of isospin breaking in the masses of the hadrons.
\end{itemize}
From this, the reference states are now postdicted at
\begin{itemize}
\item[(a)] For the $\Lambda_c(2940)$, the new mass is $2939.2\,{\rm MeV}$
  (previously: $2939.6$) and the $L$ and $H$ vertex functions are
  now $(\phi_L, \phi_H) = (0.76,-0.65)$, indicating a small
  deviation with respect to a pure $I=0$ state.
\item[(b)] For the $\Sigma_c(2800)$, the mass is $\sim 2800.6\,{\rm MeV}$
  (before: $\sim 2800\,{\rm MeV}$) and $(\phi_L, \phi_H) = (0.41,0.91)$,
  i.e. a larger deviation from the isospin symmetric limit
  when compared with the $\Lambda_c(2940)$.
\end{itemize}
That is, for the masses of the two previous molecular states
isospin symmetry breaking seems to be a perturbative
correction over the isospin symmetric limit.
But this is only true provided the mass of the molecular state is predicted
below the $L$ channel threshold: for predictions above the $L$ threshold,
which is what happens in the $D \Xi$ and $D^* \Xi$ systems,
there will be significant changes in the predicted masses.

In the particular case of the $D \Xi$ and $D^* \Xi$ molecules,
the two particles channels corresponding to the $I=0,1$, $M_I=0$
configurations are relatively far away from each other
\begin{eqnarray}
  m(D^0 \Xi^0) &=& 3179.3 \,{\rm MeV} \, , \\
  m(D^+ \Xi^-) &=& 3191.4 \,{\rm MeV} \, , \\
  \nonumber \\
  m(D^{*0} \Xi^0) &=& 3321.8 \,{\rm MeV} \, , \\
  m(D^{*+} \Xi^-) &=& 3332.0 \,{\rm MeV} \, , 
\end{eqnarray}
where the predictions of the saturation model fall in between
the two thresholds when the reference state is
the $\Lambda_c(2940)$.
In this later case, concrete calculations show that the $I=0$ and $I=1$
$\Xi D$ states we originally predicted in Table \ref{tab:molecular-baryons}
now become a pair of predominantly $\Xi^0 D^{(*)0}$ and $\Xi^- D^{(*)+}$
states, as shown in Table \ref{tab:molecular-omegac-isospin}.
For $\Xi D$ (with $\Lambda_c(2940)$ as the reference state)
we predict the masses
\begin{eqnarray}
  m(\Xi^0 D^0 (L)) &=& 3178.5\,{\rm MeV} \, , \\ 
  m(\Xi^- D^+ (H)) &=& 3190.7\,{\rm MeV} \, ,
\end{eqnarray}
where the higher energy state is relatively close to the experimental mass
($M = 3185^{+7.6}_{-1.9}\,{\rm MeV}$).
Conversely, for $J=\tfrac{3}{2}$ $D^* \Xi$ we predict now
\begin{eqnarray}
  m(\Xi^0 D^{*0} (L)) = 3320.8\,{\rm MeV} \, , \\
  m(\Xi^- D^{*+} (H)) = 3331.2\,{\rm MeV} \, .
\end{eqnarray}
Again, the heavier molecule is not far away from the experimental mass
($M = 3327.1^{+1.2}_{-1.8}\,{\rm MeV}$).
The vertex functions for the $L$ and $H$ channels, $\phi_L$ and $\phi_H$,
are also listed in Table \ref{tab:molecular-omegac-isospin},
where it is apparent that isospin is badly broken at the level of
the wave function and neither of the two states
can be interpreted as a $I=0$ or $I=1$ state.
However, when we use the $\Sigma_c(2800)$ as the reference state, which implies
more attraction, and the prediction of the $I=0$ state happens below
the $L$ threshold in the isospin symmetric limit, then the changes
in the mass after including isospin breaking in the masses
are relatively small, check Tables \ref{tab:molecular-baryons}
and \ref{tab:molecular-omegac-isospin}.

As a consequence, if the $\Omega_c(3185/3327)$ are molecular
they should appear as a double peak: (i) a peak close to the
$D^{(*)+} \Xi^-$ threshold, roughly corresponding to
what is seen in the experiment, and (ii) a second,
lighter peak close to the $D^{(*)0} \Xi^0$ threshold.
{\it Prima facie}, this seems to contradict the experimental findings,
as there is no $\Omega_c$ listed with the mass of the lighter peak.
Yet, regarding the $\Omega_c(3185)$,  Ref.~\cite{LHCb:2023rtu} states:
{\it ``A two-peak structure also describes the data well in the mass region
around 3185 MeV, hence the presence of two states in this region
can not be excluded.''}.

Unfortunately, the masses of the two-peak solution are not given,
neither it is said explicitly whether this also applies to
the $\Omega_c(3327)$.
For the later, in Table \ref{tab:molecular-omegac-isospin} we predict
that the $J=\tfrac{1}{2}$ and $\tfrac{3}{2}$ $\Xi^- D^{*+}$ peaks are
almost at the same mass, which (within the two-peak hypothesis)
might explain why the uncertainties in the $\Omega_c(3327)$ mass are
much smaller ($3327.1^{+1.2}_{-1.8}\,{\rm MeV}$) than those of
the $\Omega_c(3185)$ ($3185.1^{+7.6}_{-1.9}\,{\rm MeV}$).

Moreover, if the $\Omega_c(3185)$ and $\Omega(3327)$ were double peaks,
this factor could indeed explain their large observed widths
in~\cite{LHCb:2023rtu}.
A pure molecular explanation will result in a narrow state because
the $\Xi D \to \Xi_c \bar{K}$ transition depends on short-range dynamics
(e.g. the exchange of a light-baryon).
These dynamics are expected to be suppressed
if the wave function has a large size.
Within this scenario, the relatively large width (for a molecular state) of
the experimental peaks could be a consequence of its double peak nature.

More insight might be gained from a comparison
with the compact hadron hypothesis~\footnote{Regarding this hypothesis,
  we mention in passing that recently Ref.~\cite{Luo:2023sra} has
  proposed that the $\Omega_c(3327)$ is a compact $1D_{\frac{5}{2}}$ state,
  while Ref.~\cite{Karliner:2023okv} interprets the $\Omega_c(3185)$ and
  $\Omega_c(3327)$ as $2S_{\frac{1}{2}}$ and $2S_{\frac{3}{2}}$ states.}.
From the equal spacing rule, we expect the partners of the $\Omega_c(3185)$ and
$\Omega(3327)$ to have a similar spacing to that of the lowest mass charmed
baryon sextet, that is, $M(\Omega_c) - M(\Xi_c') \sim 125\,{\rm MeV}$ and
$M(\Xi_c') - M(\Sigma_c) \sim 125\,{\rm MeV}$.
Thus we might expect the sextets:
\begin{eqnarray}
  \Sigma_c(2940) \quad , \quad \Xi_c(3060) \quad , \quad \Omega_c(3185) \, , \\
  \Sigma_c(3075) \quad , \quad \Xi_c(3200) \quad , \quad \Omega_c(3327) \, .
\end{eqnarray}
Currently, only the $\Xi_c(3055)$ fits within the previous pattern.
However, the identification of the $\Xi_c(3055)$ as a partner of
the $\Omega_c(3185)$ is problematic in what regards the widths:
the $\Xi_c(3055)$ has a width of a few ${\rm MeV}$
($\Gamma = 7.8 \pm 1.9\,{\rm MeV}$~\cite{ParticleDataGroup:2022pth}),
while for the $\Omega_c(3185)$ it is of a few tens of ${\rm MeV}$
($\Gamma = 50 \pm 7 {}^{+10}_{-20}\,{\rm MeV}$~\cite{LHCb:2023rtu}),
a difference of one order of magnitude.
This suggests that they are not partners, though confounding factors might
exist: a compact $\Xi_c(3055)$ and $\Omega_c(3185)$ could both mix
with the nearby meson-baryon thresholds, muddling the comparison
between the two.

\begin{table*}[!ttt]
\begin{tabular}{|c|c|ccc|ccc|c|}
\hline\hline
System & % $S$ &
$J^P$ & $R_{\rm mol}$(${\Lambda_c^*}$) &
$(\phi_L, \phi_H)$ & $M_{\rm mol}$ &
$R_{\rm mol}$(${\Sigma_c^*}$) &
$(\phi_L, \phi_H)$ & $M_{\rm mol}$ & $M_{\rm cand}$ 
\\
\hline
\hline
\multirow{2}{*}{$\Xi^0 {D}^0$-$\Xi^- D^+$} % & \multirow{2}{*}{$-2$}
& \multirow{2}{*}{$\frac{1}{2}^-$} & \multirow{2}{*}{$\begin{pmatrix} 0.82 & -0.08 \\ -0.08 & 0.82 \end{pmatrix}$} & $(-0.86,0.51)$ & $3178.5^{+1.2}_{-3.0}$
& \multirow{2}{*}{$\begin{pmatrix} 1.23 & -0.12 \\ -0.12 & 1.24 \end{pmatrix}$}
& $(-0.86, 0.51)$ & $3159.7^{+2.1}_{-3.4}$ & - \\
&  & & $(0.07-0.30\,i,0.95)$ & $3190.7^{+0.8}_{-3.0}$
&  & $(0.28, 0.96)$ & $3177.6^{+2.1}_{-1.4}$ &
$3185.1^{+7.6}_{-1.9}$~\cite{LHCb:2023rtu} \\
\hline
\multirow{2}{*}{$\Xi^0 {D}^{*0}$-$\Xi^- D^{*+}$} % & \multirow{2}{*}{$-2$}
& \multirow{2}{*}{$\frac{1}{2}^-$} & \multirow{2}{*}{$\begin{pmatrix} 0.89 & -0.14 \\ -0.14 & 0.89 \end{pmatrix}$} & $(-0.86,0.51)$ & $3317.1^{+2.3}_{-3.1}$
& \multirow{2}{*}{$\begin{pmatrix} 1.35 & -0.21 \\ -0.21 & 1.36 \end{pmatrix}$}
& $(-0.79, 0.62)$ & $3289.3^{+7.0}_{-9.8}$ & - \\
&  & & $(0.07-0.30\,i,0.95)$ & $3330.3^{+2.2}_{-3.7}$
&  & $(0.45, 0.89)$ & $3316.5^{+2.4}_{-1.8}$ & - \\
\hline
\multirow{2}{*}{$\Xi^0 {D}^{*0}$-$\Xi^- D^{*+}$} % & \multirow{2}{*}{$-2$}
& \multirow{2}{*}{$\frac{3}{2}^-$} & \multirow{2}{*}{$\begin{pmatrix} 0.81 & -0.06 \\ -0.06 & 0.82 \end{pmatrix}$} & $(-0.97,0.26)$ & $3320.8^{+0.8}_{-3.3}$
& \multirow{2}{*}{$\begin{pmatrix} 1.23 & -0.09 \\ -0.09 & 1.24 \end{pmatrix}$}
& $(-0.89, 0.45)$ & $3303.9^{+0.4}_{-0.9}$ & - \\
&  & & $(0.00-0.16\,i,0.99)$ & $3331.2^{+0.7}_{-3.2}$
&  & $(0.27, 0.96)$ & $3317.8^{+2.2}_{-1.7}$ & $3327.1^{+1.2}_{-1.8}$~\cite{LHCb:2023rtu} \\
\hline \hline
\end{tabular}
\caption{Predictions for the $\Omega_c$ molecular baryons when isospin breaking
  effects in the masses of the $\Xi^0 D^{0(*)}$ and $\Xi^- D^{+(*)}$ are taken
  into account. ``System'' refers to the particular $\Xi D^{(*)}$ molecule
  under consideration, $J^P$ to its spin and parity, $R_{\rm mol}$ is
  the relative strength of the contact-range interaction as defined
  in Eq.~(\ref{eq:Rmol-AB}), $(\phi_L, \phi_H)$ the vertex function
  for the lower and higher mass channels , $M_{\rm mol}$ the mass
  of the predicted state and $M_{\rm cand}$ the mass of
  the $\Omega_c$ candidate states.
  The uncertainties in $M_{\rm mol}$ come from varying the scalar meson mass
  in the $(400-550)\,{\rm MeV}$ range.
  All masses are in units of ${\rm MeV}$.
}
\label{tab:molecular-omegac-isospin}
\end{table*}

\section{Conclusions}

We have considered the spectroscopy of charmed meson and octet baryon molecules
within a phenomenological model.
This model is a contact-range theory in which the couplings are saturated by
the exchange of the light scalar and vector mesons ($\sigma$, $\rho$,
$\omega$, $K^*$ and $\phi$).
The choice of a contact-range interaction is motivated by the difference
in scales between the range of light-meson exchange (short-range)
and the size of the molecular states predicted (long-range).
The saturation of the couplings exploits their RG evolution to combine
the contributions from light-mesons with different masses.
The couplings are determined up to a proportionality constant that has
to be calibrated by reproducing a given reference state, i.e. a known
state with a plausible molecular interpretation.
For this we use the $\Lambda_c(2940)$ (as an $I=0$, $J=\tfrac{3}{2}$
$N D^*$ molecule) and the $\Sigma_c(2800)$ ($I=1$, $J=\tfrac{1}{2}$
$ND$ molecule).
Each reference state leads to quantitative differences in the
charmed baryon and anticharmed pentaquark spectra.

Among the molecular charmed baryons we predict,
there are $\Sigma D$ and $\Lambda D_s^*$ bound states that might correspond
with the $\Xi_c(3055)$ and $\Xi_c(3123)$ baryons.
Yet, the more interesting result might be the prediction of $\Xi D$ and
$\Xi D^*$ bound states with masses matching those of the recently
observed $\Omega_c(3185)$ and $\Omega_c(3327)$.
For this molecular interpretation to be valid it would be required that
the $\Omega_c(3185)$ is composed of two narrow peaks with a mass
difference of about $10\,{\rm MeV}$ (i.e. the gap between
the $\Xi^- D^{+}$ and $\Xi^0 D^{0}$ thresholds).
It is noteworthy that the $\Omega_c(3185)$ indeed accepts a two peak
description~\cite{LHCb:2023rtu}, though the masses of each of
the peaks is not mentioned.
For the $\Omega_c(3327)$ the situation might be more complex because
the two spin configurations ($J=\tfrac{1}{2}$ and $\tfrac{3}{2}$) of
the $\Xi D^*$ system bind, meaning that there could be up to four
peaks (though this might depend on the magnitude of
the isospin splitting).
Yet, the $J=\tfrac{1}{2}$ and $\tfrac{3}{2}$ $\Xi^- D^{*+}$ bound states
are predicted about the same mass, representing a simplification
with respect to the four peak scenario.
In this latter case,
if the $\Omega_c(3327)$ turns out to contain two nearby peaks
with a mass difference smaller than the $\Xi^- D^{*+}$ and $\Xi^0 D^{0}$
thresholds gap, this would support a molecular interpretation.

Finally, we predict a few molecular anticharmed pentaquarks.
In this case there are no experimental candidates and the only comparison
left is with other theoretical models~\cite{Gignoux:1987cn,Lipkin:1987sk,Hofmann:2005sw,Yalikun:2021dpk}, which in general do agree on the
qualitative features of the molecular spectrum
(for instance, the possibility of
$N \bar{D}_s^{(*)}$~\cite{Gignoux:1987cn,Lipkin:1987sk,Hofmann:2005sw} or
$\Sigma \bar{D}^{(*)}$~\cite{Yalikun:2021dpk} states).
Yet, there is experimental information about the $I=0$ $N \bar{D}$ interaction
from the ALICE collaboration~\cite{ALICE:2022enj}:
its inverse scattering length.
This datum is reproduced by our RG saturation model independently of
the input ($\Lambda_c(2940)$ or $\Sigma_c(2800)$).

\section*{Acknowledgments}

This work is partly supported by the National Natural Science Foundation
of China under Grants No. 11735003. No. 11975041, No. 11835015,  No. 12047503
and No. 12125507, the Chinese Academy of Sciences under Grant No. XDB34030000,
the China Postdoctoral Science Foundation under Grant No. 2022M713229,
the Fundamental Research Funds for the Central Universities and
the Thousand Talents Plan for Young Professionals.
M.P.V. would also like to thank the IJCLab of Orsay, where part of
this work has been done, for its long-term hospitality.

\appendix
\section{Light-meson couplings}
\label{app:couplings}

Here we explore in more detail our choice of couplings for the light baryons.
We begin with the vector meson couplings, which are derived
from the mixing with the electromagnetic current.
We continue with the scalar couplings, whose choice requires a more
careful comparison with molecular predictions in a few system.
This leads to the conclusion that this coupling is weaker for the $\Lambda$
than for the other octet baryons.

\subsection{Vector couplings}

For the vector couplings, we determined them from the fact that the neutral
vector mesons can mix with the photon current (because they have the
same charge and quantum numbers $J^{PC} = 1^{--}$), i.e. from vector meson
dominance~~\cite{Sakurai:1960ju,Kawarabayashi:1966kd,Riazuddin:1966sw}.
For this we consider the non-relativistic interaction between a hadron $h$
and a vector meson $V$ as given by the Lagrangian
\begin{eqnarray}
  \mathcal{L}_{h h V} = g_V\,h^{\dagger}\,\left[\,\partial_0 V_0
    + \frac{\kappa_V}{2 M}\,\epsilon_{ijk}\,\hat{S}_i\,\partial_j\,V_k
    \right]\,h \, ,
\end{eqnarray}
with $g_V$ and $\kappa_V$ the electric- and magnetic-like couplings,
$\hat{S}_i$ the $i=1,2,3$ spatial component of the reduced spin
operator $\hat{\vec{S}} = \vec{S} / S$ (with $S$ the spin of
hadron $h$) and $M$ a scaling mass, for which we choose
the nucleon mass.
For simplicity we have not indicated the isospin or flavor indices
explicitly.
Next we make the substitutions
\begin{eqnarray}
  \rho^0_{\mu} &\to& \rho^0_{\mu} + \frac{1}{2}\frac{e}{g}\,A_\mu \, , \\
  \omega_{\mu} &\to& \omega_{\mu} + \frac{1}{6}\frac{e}{g}\,A_\mu \, , \\
  \phi_{\mu} &\to& \phi_{\mu} - \frac{1}{3 \sqrt{2}}\frac{e}{g}\,A_\mu \, ,
\end{eqnarray}
which depend on whether the vector meson is a neutral $\rho$ ($\rho^0$),
an $\omega$ or a $\phi$, $\mu$ is a Lorentz index and
$A_{\mu}$ the photon field; $e$ is the proton charge and
$g = m_V / 2 f_{\pi} \simeq 2.9$ the universal vector meson coupling
(in the $f_{\pi} = 132\,{\rm MeV}$ normalization).
We match the $A_{\mu}$ piece of the previous substitution to
the non-relativistic electromagnetic Lagrangian
for the light-quark content of the hadron $H$
\begin{eqnarray}
  \mathcal{L}_{h h \gamma} = e\,h^{\dagger}\,\left[\,Q_L \partial_0\,A_0
    + \frac{\mu_L}{2 M}\,\epsilon_{ijk}\,\hat{S}_i\,\partial_j\,A_k
    \right]\,h \, , \nonumber \\
\end{eqnarray}
where $Q_L$ is the charge of the light quarks within
the hadron $h$ (in units of $e$) and $\mu_L$ its magnetic moments
in units of $e / (2 M)$ (or nuclear magnetons, if $M$ is chosen
to be the nucleon mass).
Of course, if isospin or flavor are explicitly consider,
$Q_L$ and $\mu_L$ will become matrices.
The $g_V$ couplings depend on the charges of the isospin components
of the hadrons, while $\kappa_V$ on their (light) magnetic moments.
By using the quark model calculation of the magnetic moments of the octet
baryons and the part of the charmed meson magnetic moments
that come from the light-quarks, we arrive at the $\kappa_V$
couplings of Table \ref{tab:couplings}.

\subsection{Scalar coupling}

Here we explore in more detail the couplings of the scalar meson
in the strange sector for the light baryons and charmed mesons.
Our baseline scenario is that this couplings is given by
$g_{Sqq} = 3.4$ for $q = u, d, s$, as derived from the linear sigma
model~~\cite{GellMann:1960np}, the quark model~\cite{Riska:2000gd} and
the additional assumption that the coupling to the $s$ quark
is similar to the $u$ and $d$ quarks.
We will decide whether this baseline value requires corrections or not
by calculating the spectra of a few two-hadron systems and comparing
with experimental information or other theoretical models.

For the coupling of the scalar meson to the light baryons,
we calculate a few two light baryon systems using
the two-nucleon $^1S_0$ virtual state as
the reference state (or, equivalently, by using the $^1S_0$ scattering
length as input, $a_0({}^1S_0) = -23.7\,{\rm fm}$).
First, we notice that in terms of SU(3) symmetry, the two-nucleon $^1S_0$
configuration and a series of other configurations:
\begin{eqnarray}
  | NN({}^1S_0, I=1) \rangle &=& | 27 \rangle \, , \\
  | \Sigma N({}^1S_0, I=\frac{3}{2}) \rangle &=& | 27 \rangle \, , \\
  | \Sigma \Sigma({}^1S_0, I=2) \rangle &=& | 27 \rangle \, , \\
  | \Xi \Sigma ({}^1S_0, I=\frac{3}{2}) \rangle &=& | 27 \rangle \, , \\
  | \Xi \Xi ({}^1S_0, I=1) \rangle &=& | 27 \rangle \, , 
\end{eqnarray}
are all in the $27$-plet SU(3)-flavor representation, from which
the potential should be the same in the flavor-symmetric limit.
Indeed, all of these systems happen to show large scattering lengths
that are in a few cases positive (indicating a bound state).
Following~\cite{Peng:2021hkr} we use a softer cutoff
in the light sector, $\Lambda = 0.5\,{\rm GeV}$,
in which case saturation yields
\begin{eqnarray}
  B_2(\Sigma N, I=\tfrac{3}{2}) &=& 1.1 \, {\rm MeV} \, , \\
  B_2(\Sigma \Sigma, I=2) &=& 1.6 \, (0-0.01)\, {\rm MeV} \, , \\
  B_2(\Xi \Sigma, I=\tfrac{3}{2}) &=& 1.0 \,(0.58-0.19)\, {\rm MeV} \, , \\
  B_2(\Xi \Xi, I=1) &=& 2.1 \,(0.40-1.0) {\rm MeV} \, , 
\end{eqnarray}
where the values in parentheses correspond to chiral EFT results
when terms up to order $Q^2$ are included
in the potential~\cite{Haidenbauer:2014rna},
where we notice that (i) for the $^1S_0$ $\Sigma N$ no bound state is
predicted in~\cite{Haidenbauer:2014rna}, though there is considerable
attraction if we look at the scattering length and (ii) that
the order $Q^0$ results would be more similar to our
$\Xi \Sigma$ ($(2.23-6.18)\,{\rm MeV}$ in~\cite{Haidenbauer:2014rna}
versus our $1.0\,{\rm MeV}$ result) and $\Xi \Xi$ predictions
($(2.56-7.27)\,{\rm MeV}$ in~\cite{Haidenbauer:2014rna}
versus $2.1\,{\rm MeV}$).
Taking into account that we are not considering exchange of pseudoscalar mesons,
which lead to less attraction in the strangeness $S=-1$ and $-2$ system
relative to the singlet, the results we obtain are sensible.
We could have also compared the scattering lengths,
in which case we would have had
\begin{eqnarray}
  a_0(\Sigma N) &=& 6.4\,{\rm fm} \, , \\
  a_0(\Sigma \Sigma) &=& 5.2\,(60.6-(-286.0)){\rm fm} \, , \\
  a_0(\Xi \Sigma) &=& 6.2\,(8.4-13.8)\,{\rm fm} \, , \\
  a_0(\Xi \Xi) &=& 4.4\,(9.7-6.5)\,{\rm fm} \, , 
\end{eqnarray}
where results in parentheses are again from Ref.~\cite{Haidenbauer:2014rna}.

Alternatively, we can compare instead to the lattice QCD results of
Ref.~\cite{NPLQCD:2020lxg}, which after extrapolation to
the physical pion mass lead to a bound $^1S_0$ system
with binding energy
\begin{eqnarray}
  B_2^{\rm lin (quad)}(NN, I=0) &=& 6.4^{+6.3}_{-6.5}\,(9.9^{+4.6}_{-4.5})\,{\rm MeV}
  \, , \nonumber \\
\end{eqnarray}
depending on whether a linear or quadratic (in parentheses)
extrapolation is used to reach $m_{\pi} = 138\,{\rm MeV}$.
By using the linear extrapolation as input we obtain
\begin{eqnarray}
  B_2(\Sigma N, I=\tfrac{3}{2}) &=& 14.9 \,(8.4^{+7.8}_{-6.6}) {\rm MeV} \, , \\
  B_2(\Sigma \Sigma, I=2) &=& 15.4 \,(1.0 \pm 6.1) {\rm MeV} \, , \\
  B_2(\Xi \Sigma, I=\tfrac{3}{2}) &=& 12.8 \,(5.9^{+5.7}_{-5.8}) {\rm MeV} \, , \\
  B_2(\Xi \Xi, I=1) &=& 16.2 \,(9.6^{+4.5}_{-4.7}) {\rm MeV} \, , 
\end{eqnarray}
while if we use the quadratic extrapolation as input
\begin{eqnarray}
  B_2(\Sigma N, I=\tfrac{3}{2}) &=& 20.1 \,(11.5^{+5.7}_{-4.8})\, {\rm MeV} \, , \\
  B_2(\Sigma \Sigma, I=2) &=& 20.4 \,(5.8^{+4.2}_{-4.3})\, {\rm MeV} \, , \\
  B_2(\Xi \Sigma, I=\tfrac{3}{2}) &=& 17.2 \,(9.5^{+3.8}_{-4.0})\,
  {\rm MeV} \, , \\
  B_2(\Xi \Xi, I=1) &=& 21.1 \,(12.4^{+3.0}_{-3.1})\, {\rm MeV} \, ,
\end{eqnarray}
where the results in parentheses are from Ref.~\cite{NPLQCD:2020lxg}.
In this case our predictions tend to bind more
than the extrapolated lattice results.
The point is though that the naive choice of couplings works (within reason)
in this particular case, and thus we do not modify it
for the $N$, $\Sigma$ and $\Xi$ baryons.

Yet, for the $\Lambda$ baryon we actually have to modify its coupling to the
scalar meson in order to reproduce current theoretical estimations of
the $\Lambda N$ and $\Lambda \Lambda$ scattering length.
If we use $g_{\sigma \Lambda \Lambda} = g_{\sigma N N}$ (and the $^1S_0$ virtual
state as a reference state), in general we find excessive attraction,
where the scattering lengths are
\begin{eqnarray}
  a_0(\Lambda N, {}^1S_0) &=& 53.1\,{\rm fm} \, , \\
  a_0(\Lambda N, {}^1S_0) &=& 346.8\,{\rm fm} \, , \\
  a_0(\Lambda \Lambda) &=& 16.2\,{\rm fm} \, , 
\end{eqnarray}
where the positive scattering lengths indicate the existence of
bound states, in disagreement with other theoretical models.
In contrast, for $g_{\sigma \Lambda \Lambda} = (3/4)\,g_{\sigma N N}$ we obtain
\begin{eqnarray}
  a_0(\Lambda N, {}^1S_0) &=& -3.1\,{\rm fm} \, , \\
  a_0(\Lambda N, {}^1S_0) &=& -2.9\,{\rm fm} \, , \\
  a_0(\Lambda \Lambda) &=& -1.3\,{\rm fm} \, , 
\end{eqnarray}
which compare well (though not perfectly) with other models:
(i) for the $\Lambda N$ case, we have
$a_0(\Lambda N, {}^1S_0) = -2.9$, $-2.6$ and $-2.6\,{\rm fm}$ and
$a_0(\Lambda N, {}^3S_1) = -1.7$, $-1.7$ and $-1.7\,{\rm fm}$
with chiral potentials at the next-to-leading order
(NLO)~\cite{Haidenbauer:2013oca},
the J\"ulich 04 model~\cite{Haidenbauer:2005zh} and
the Nijmegen soft core potential~\cite{Rijken:1998yy},
respectively, while (ii) for the $\Lambda \Lambda$ case,
$a_0(\Lambda \Lambda) = -(0.33-0.85)\,{\rm fm}$ in chiral NLO~\cite{Haidenbauer:2015zqb}, $a_0(\Lambda \Lambda) = -0.81 \pm 0.23 {}^{+0.0}_{-0.13}\,{\rm fm}$
in the lattice~\cite{HALQCD:2019wsz}.
Even though it is possible to further fine-tune the parameters to match
better the results of other models, we consider that the current
change ($g_{S \Lambda \Lambda} = (3/4) g_{S NN}$) is enough
for our purposes.

The charmed meson case is simpler. Here we will use the recent
lattice QCD prediction of $J^{PC} = 0^{++}$ $D\bar{D}$ and
$D_s\bar{D}_s$ bound states~\cite{Prelovsek:2020eiw}
as pseudodata.
We notice in passing that the $X(3960)$~\cite{LHCb:2022vsv}
is interpreted as a $D_s \bar{D}_s$ molecular state
too~\cite{Ji:2022uie,Abreu:2023rye},
where the location of the pole (bound or virtual)
is usually not far away from the lattice result.
For this prediction, we use the $X(3872)$ as the reference state (interpreted
as a $I=0$, $J^{PC} = 1^{++}$ $D^*\bar{D}$ bound state) and a cutoff of
$\Lambda = 1.0\,{\rm GeV}$ as in~\cite{Peng:2021hkr}.
The binding energies of the $D\bar{D}$ state is calculated to be
\begin{eqnarray}
  B_2(D \bar{D}) &=& 4.0^{+5.0}_{-3.7} \, {\rm MeV} \, , 
\end{eqnarray}
and we will use it as input in our calculations.
For the $D_s \bar{D}_s$ state there are two calculations, a single channel one
in which a bound state is found
\begin{eqnarray}
  B_2^{\rm SC}(D_s \bar{D}_s) &=& 6.2^{+2.0}_{-3.8} \, {\rm MeV} \, , 
\end{eqnarray}
and a coupled channel one, in which we have a resonance instead with energy
\begin{eqnarray}
  E_2^{\rm CC}(D_s \bar{D}_s) &=& -0.2^{+0.17}_{-4.9} - \frac{i}{2}\,(0.27^{+2.5}_{-0.15}) \, {\rm MeV} \, , \nonumber \\ 
\end{eqnarray}
where $E_2$ is the energy of the state with respect to the $D_s\bar{D}_s$
threshold.
If we assume $g_S' = g_S$, we will predict this state to be at
\begin{eqnarray}
  B_2^{\rm SC}(D_s \bar{D}_s) &=& (1.0)^V \, {\rm MeV} \, , \\
  E_2^{\rm CC}(D_s \bar{D}_s) &=& (-2.4 - \frac{i}{2}\,1.5)^V \, {\rm MeV} \, , 
\end{eqnarray}
with both solutions corresponding to a virtual state (where in the coupled
channel case this specifically means a pole in the (I,II) Riemann sheet).
Even though outside the error bands of the lattice predictions,
these two results are still in line with them.
From this point of view, it might not be necessary to tweak the scalar coupling.
If we take $g_{S D_s D_s} = 1.15\,g_{S D D}$ instead, we will predict
\begin{eqnarray}
  B_2^{\rm SC}(D_s \bar{D}_s) &=& 1.5 \, {\rm MeV} \, , \\
  E_2^{\rm CC}(D_s \bar{D}_s) &=& (-0.25 - \frac{i}{2}\,0.42) \, {\rm MeV} \, , 
\end{eqnarray}
where the single channel calculation is now a bound state, in agreement
with Ref.~\cite{Prelovsek:2020eiw}, and the coupled channel calculation
a resonance in the (II,I) Riemann sheet (we notice that in this case,
Ref.~\cite{Prelovsek:2020eiw} finds that this state is in the (II,I) sheet
in $70\%$ of the bootstrap samples and in (I,II) in the rest).
However, even though this change improves the agreement with lattice,
we do not consider that it is necessary to include it (the improvement
is marginal) and will opt instead for the more simple
$g_{S D_s D_s} = g_{S D D}$ choice.
Finally, we also notice that the reproduction of the $Z_{cs}(3985)$ as a
$D^* \bar{D}_s$-$D \bar{D}_s^*$ molecule from the $Z_c(3900)$
($D^* \bar{D}$) also requires $g_{S D_s D_s} \geq g_{S D D}$~\cite{Yan:2021tcp}. 

%\bibliography{refs.bib}

\end{document}